\documentclass[twocolumn]{myaastex631} 
\usepackage{graphicx,url}
\usepackage{amsmath,amssymb}
\usepackage{hyperref}
\hypersetup{colorlinks,
	linkcolor=blue, 
	urlcolor=magenta, 
	anchorcolor=blue,
	citecolor=blue
}
\usepackage{bm}
\usepackage{xcolor}
\usepackage{mathrsfs}
\usepackage{natbib}
\usepackage{soul} 

\definecolor{greenW}{rgb}{0.0, 0.55, 0.1}
\definecolor{orangeW}{rgb}{1.0, 0.5, 0.05}

\shorttitle{ Long Short-Term Memory networks}
\shortauthors{Cui et al.}
\begin{document}

\title{Advancing Cosmological Parameter Estimation and Hubble Parameter Reconstruction with Long Short-Term Memory and Efficient-Kolmogorov-Arnold Networks}

\author{Jiaxing Cui}
\affiliation{Department of Physics, and Collaborative Innovation Center for Quantum Effects and Applications, Hunan Normal University, Changsha 410081, China;}
\affiliation{College of Information Science and Engineering, Hunan Normal University, Changsha, Hunan 410081, People's Republic of China;}

\author{Marek Biesiada}
\affiliation{National Centre for Nuclear Research, Pasteura 7, 02-093 Warsaw, Poland;}

\author{Ao Liu}
\affiliation{Department of Physics, and Collaborative Innovation Center for Quantum Effects and Applications, Hunan Normal University, Changsha 410081, China;}
\affiliation{College of Information Science and Engineering, Hunan Normal University, Changsha, Hunan 410081, People's Republic of China;}

\author{Cuihong Wen$^{\ast}$}
\affiliation{Department of Physics, and Collaborative Innovation Center for Quantum Effects and Applications, Hunan Normal University, Changsha 410081, China;}
\affiliation{College of Information Science and Engineering, Hunan Normal University, Changsha, Hunan 410081, People's Republic of China;}

\author{Tonghua Liu$^{\star}$}
\affiliation{School of Physics and Optoelectronic Engineering, Yangtze University, Jingzhou, 434023, China;}

\author{Jieci Wang$^{\dagger}$}
\affiliation{Department of Physics, and Collaborative Innovation Center for Quantum Effects and Applications, Hunan Normal University, Changsha 410081, China;}
\email{$^{\ast}$cuihongwen@hunnu.edu.cn;\\$^{\dagger}$liutongh@yangtzeu.edu.cn;\\$^{\dagger}$jcwang@hunnu.edu.cn}

\begin{abstract}
In this work, we propose a novel approach for cosmological parameter estimation and Hubble parameter reconstruction using Long Short-Term Memory (LSTM) networks and Efficient-Kolmogorov-Arnold Networks (Ef-KAN). LSTM networks are employed to extract features from observational data, enabling accurate parameter inference and posterior distribution estimation without relying on solvable likelihood functions. This method achieves performance comparable to traditional Markov Chain Monte Carlo (MCMC) techniques, offering a computationally efficient alternative for high-dimensional parameter spaces.  By sampling from the reconstructed data and comparing it with mock data, our designed LSTM constraint procedure demonstrates the superior performance of this method in terms of constraint accuracy, and effectively captures the degeneracies and correlations between the cosmological parameters. Additionally, the Ef-KAN model is introduced to reconstruct the Hubble parameter $H(z)$ from both observational and mock data. Ef-KAN is entirely data-driven  approach, free from prior assumptions, and demonstrates superior capability in modeling complex, non-linear data distributions.  We validate the Ef-KAN method by reconstructing the Hubble parameter, demonstrating that  $H(z)$ can be reconstructed with high accuracy. By combining LSTM and Ef-KAN, we provide a robust framework for cosmological parameter inference and Hubble parameter reconstruction, paving the way for future research in cosmology, especially when dealing with complex datasets and high-dimensional parameter spaces.
\end{abstract}
\keywords{cosmological parameters --- cosmology: observations --- methods: data analysis --- methods: computational methods --- methods: neural networks}

\section{Introduction}
In recent years, significant advancements in the exploration of our universe have been achieved through the integration of diverse observational datasets, including Type Ia supernovae  \citep{2007ApJ...659...98R,Scolnic2018}, Cosmic Microwave Background \citep{Aghanim2014,Aghanim2018}, and large-scale structures \citep{Chuang2017,Pan2019}, among others.  Based on astronomical observation data and theoretical models, Bayesian inference is a commonly used method to infer the posterior distribution of model parameters. Traditional parameter estimation methodologies primarily depend on the precise computation of likelihood functions. However, the increasing complexity of observational data and the high-dimensionality of parameter spaces often render these likelihood functions computationally intensive or even intractable \citep{Weyant2013}. In recent years, likelihood-free inference techniques have garnered considerable attention as a solution to these challenges. These techniques circumvent the direct computation of likelihood functions, with Approximate Bayesian Computation (ABC; \citet{Marjoram2003,Bonassi2015}) being one of the pioneering approaches. The ABC  method approximates the posterior distribution by evaluating the distances between simulated and observed data. However, such method requires a substantial number of simulations, and its computational cost increases exponentially with both model complexity and parameter dimensionality, thereby limiting its applicability in high-dimensional spaces \citep{Alsing2019}. To overcome these limitations, neural network-based likelihood-free inference methods have been developed most recently. For instance, \citet{Wang2020ecopann} employed artificial neural networks (ANNs) for cosmological parameter inference. Techniques such as Masked Autoregressive Flows (MAF; \citet{Papamakarios2017}) and Mixture Density Networks (MDN; \citet{Bishop1994}) have been widely utilized to model conditional probability distributions ($p(d|\theta) $). Building on these advancements, recent work has used neural networks to infer cosmological parameters from observational data \citep{WangY2020,Chen2022,Wang2023}. These methods excel in processing high-dimensional data and complex probability distributions, opening new pathways for cosmological research.

On the other hand, ANNs, as a powerful non-parametric method, have demonstrated significant potential in cosmological data analysis due to their robust nonlinear modeling capabilities and minimal reliance on assumptions about data distributions \citep{Hornik1991,Cybenko1989}. ANNs excel in handling high-dimensional data and have been successfully applied to various cosmological problems, including the reconstruction of the Hubble parameter function \citep{Wang2020reann} and the analysis of OHD.
Despite these advancements, ANN models continue to exhibit certain limitations. They often struggle to accurately model non-Gaussian distributions or handle strong parameter covariances \citep{Papamakarios2016,Charnock2018}, and their performance in processing time series data is frequently hindered by difficulties in capturing long-range dependencies and sequential features. These challenges are particularly pronounced in the analysis of OHD. Additionally, the incompleteness of existing observational datasets and the reliance of many methods on specific cosmological model assumptions further complicate the reconstruction of the Hubble parameter $H(z)$. See recent references \citep{Wang2020reann,Zhang2024,Chen2024} for more  works about ANN in cosmological data analysis.

In parallel, Gaussian Processes (GP) have emerged as a widely used non-parametric method in cosmological research, valued for their model-independent nature \citep{Seikel2012,2012PhRvD..85l3530S,Montiel2014,Cai2016,GomezValent2018,2018MNRAS.478.3640M,2019ApJ...886L..23L,2022ApJ...939...37L,2025ApJ...981L..24L}. However, GP methods typically assume that observational errors follow a Gaussian distribution \citep{Sun2021}, an assumption that often does not hold for real-world data. Non-Gaussian observational errors compromise GP robustness and reconstruction accuracy \citep{Zhou:2019}. Consequently, the applicability of GP methods must be carefully evaluated, particularly in cases where data errors exhibit non-Gaussian characteristics.
These limitations highlight the necessity of developing complementary approaches to ensure reliable and precise reconstructions of cosmological parameters. While ANNs and GP demonstrate strengths, their challenges necessitate hybrid approaches combining their advantages. Future research should prioritize technique integration, performance optimization, and limitation breakthroughs to advance cosmological data analysis. These efforts are vital for achieving accurate reconstructions with complex observational datasets.

In this study, we present a novel approach based on Long Short-Term Memory (LSTM) networks to address the challenge of likelihood-free inference for cosmological parameters using OHD. Unlike traditional ANNs, LSTM networks are particularly effective in identifying long-term dependencies and sequential patterns within data  \citep{Hochreiter1997}, which makes them highly suitable for applications in cosmology. Additionally, we introduce the Efficient-Kolmogorov-Arnold Networks (Ef-KAN) model \citep{Liu2024}, which approximates unknown data distributions by integrating multiple simple components into a nonlinear transformation. Compared to conventional neural network models, the Ef-KAN demonstrates superior capability in modeling complex and highly non-linear data distributions, thereby offering enhanced potential for accurate parameter reconstruction. We further illustrate the efficacy of LSTM networks in constraining the Hubble constant and other cosmological parameters from observational data. Moreover, by applying LSTM model to the reconstruction tasks within the Ef-KAN framework, we validate the exceptional performance of the Ef-KAN model in reconstructing the $H(z)$.

This work is organized as follows: Section \ref{sec:methodology} introduces the foundational principles of LSTM networks and the Ef-KAN model. Section \ref{data} details the data used in this study, including the acquisition and preprocessing methods for OHD and simulated datasets. Section \ref{LSTMINFERENCE} begins with a brief overview of the Markov Chain Monte Carlo (MCMC) method, compared with LSTM-based parameter constraint process. This section also explores the optimization of LSTM hyper-parameters, data preprocessing techniques, and noise incorporation strategies during LSTM training. Experimental results demonstrating the application of LSTM to OHD are also presented. Section \ref{sec:reconstruct_Hz} outlines the methodology for reconstructing the $H(z)$ using mock data and subsequently demonstrates the use of LSTM for parameter estimation based on the reconstructed $H(z)$. Finally, Section \ref{sec:discussion} provides a summary of the main findings of this study and suggests possible avenues for future research.

\section{Method}\label{sec:methodology}
In this section, we present the methodologies employed in this study, namely the LSTM networks and the Ef-KAN model, detailing their theoretical foundations and applications.
\subsection{Long Short-Term Memory networks}\label{2.1}
\begin{figure*}
\begin{center}
\setlength{\abovecaptionskip}{1pt}
\includegraphics[width=0.8\textwidth]{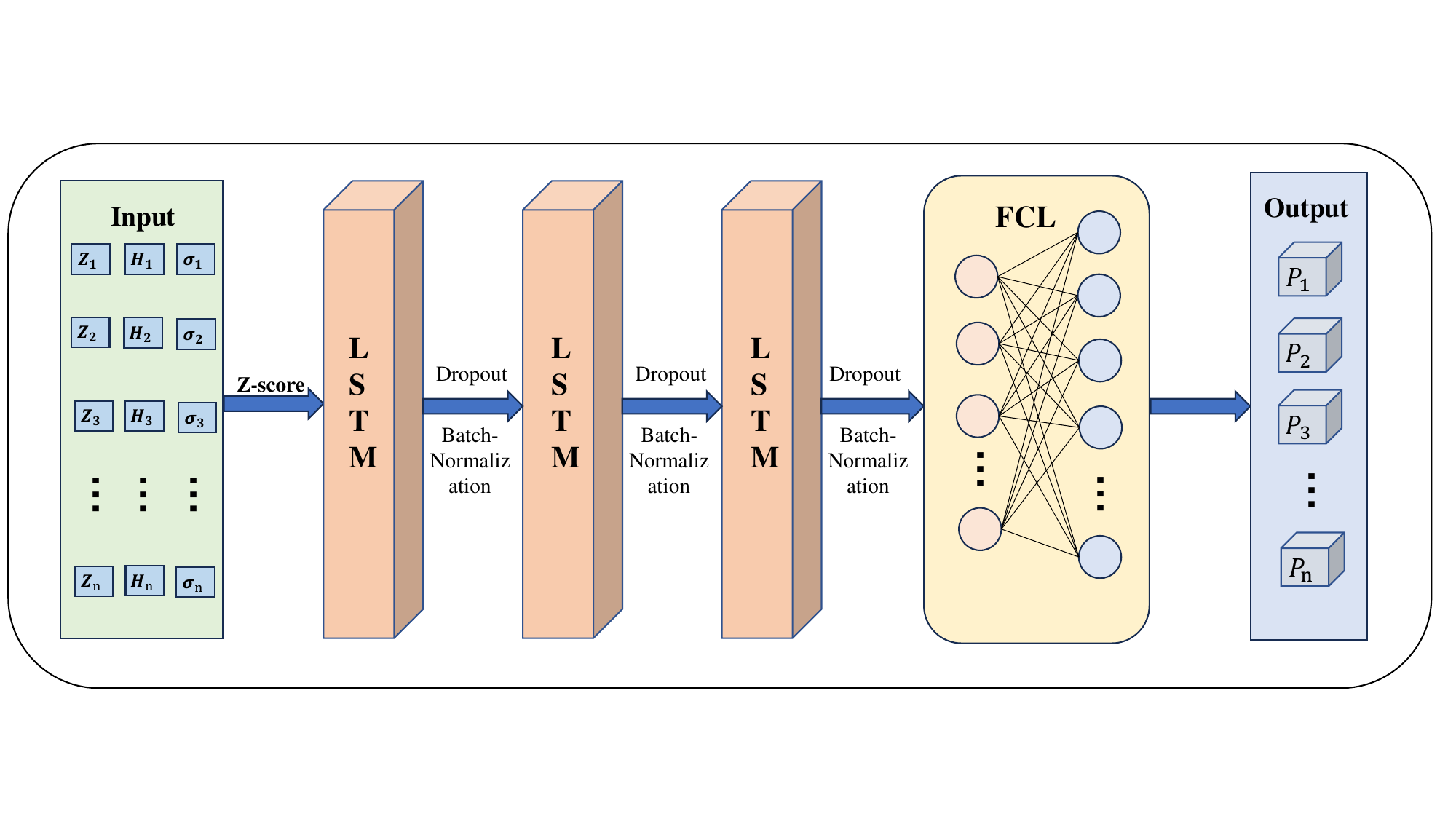}
\caption{The LSTM model introduced in this study is illustrated schematically in the figure. The leftmost box represents the input layer, which receives input data derived from the OHD. The central box, labeled “LSTM" encapsulates the LSTM modules, each comprising 256 neurons. These modules are sequentially connected to a dropout layer and a batch normalization layer to enhance generalization and stabilize training. The final Batch-Normalization layer is linked to a fully connected layer, which serves as the output layer for generating the estimated parameters. This structured design ensures robust feature extraction and effective parameter inference from the input data.}\label{fig:nn_model}
\setlength{\belowcaptionskip}{1pt}
\end{center}
\end{figure*}

The LSTM network is a specialized variant of Recurrent Neural Networks (RNNs) introduced by Hochreiter and Schmidhuber \citep{Hochreiter1997}, which has been widely adopted for processing time series data due to its ability to capture temporal dependencies. Before delving into the specifics of LSTM, it is essential to first understand the foundational principles of RNNs.

RNNs are specialized neural networks designed for the processing of sequential data, allowing information to flow through the network in a temporal manner. Owing to their specialized architecture, RNNs demonstrate strong capabilities in handling tasks that involve temporal dependencies, for example, in tasks like speech recognition and natural language understanding, and time series analysis. However, the traditional RNNs architecture struggles with modeling long-term dependencies due to the vanishing gradient issue during backpropagation \citep{Hochreiter1998}. To overcome this limitation, LSTM networks were introduced as a more robust solution for sequence modeling \citep{Hochreiter1997,Gers2000,Graves2013,Jozefowicz2015}.

While LSTM networks share a structural similarity with RNNs, their individual units are significantly more complex. Each LSTM unit comprises a memory cell and three gating mechanisms: the input gate, the forget gate, and the output gate. These gates regulate the flow of information within the unit, determining what information to retain, discard, or output. This architecture enables LSTM networks to effectively manage long-term dependencies in sequential data. The memory cell, which stores long-term information, works in conjunction with the gates to control the information flow, allowing the network to learn and maintain long-range relationships more effectively than standard RNNs.

At each time step $t$, the input $x_t$ is read by the network and contributes to the processing of future states, and the output $h_t$, along with the gate activations $i_t$ and $o_t$, is computed. Additionally, the candidate memory state  $\tilde{C}_t$ is updated conditioned on the previous hidden state $h_{t-1}$. With the current input $x_t$ and the preceding hidden state $h_{t-1}$, the following equations govern the computations within the LSTM unit:
\begin{eqnarray}\nonumber
f_t &=& \sigma \left( W_f \cdot [h_{t-1}, x_t] + b_f \right),\\\nonumber
i_t& =& \sigma \left( W_i \cdot [h_{t-1}, x_t] + b_i \right),\\\nonumber
o_t &=& \sigma \left( W_o \cdot [h_{t-1}, x_t] + b_o \right),\\
\tilde{C}_t& =& \tanh \left( W_C \cdot [h_{t-1}, x_t] + b_C \right).
\end{eqnarray}
In the equations above, the subscript $t$ indicates the current time step, while $\sigma$ and $\tanh$ denote the sigmoid and hyperbolic tangent activation functions, respectively. The weight matrices $W_i$, $W_f$, and $W_c$ are linked to the input gate, forget gate, and cell state update. Correspondingly, the bias terms $b_i$, $b_f$, $b_o$, and $b_c$ are associated with the respective gates and the cell state. Notably, these weights and biases are consistent and shared across all time steps. The resulting gate outputs and intermediate values are then employed to update both the cell state $C_t$ and the hidden state $h_t$, allowing the model to effectively preserve and transmit information throughout long sequences,
\begin{eqnarray}
C_t &= &f_t \odot C_{t-1} + i_t \odot \tilde{C}_t,\nonumber\\
h_t &=& o_t \odot \tanh(C_t).
\end{eqnarray}
The operator $\odot$ indicates element-wise multiplication between two quantities. The initial conditions are defined as $C_0 = 0$ for the memory state and $h_0 = 0$ for the hidden state. In LSTM networks, the forget gate controls whether information from the previous time step's memory unit is retained or discarded, where the input gate is responsible for incorporating and updating incoming information. At the same time, the output gate regulates the flow of information to the output of the current state. This well-defined gating mechanism enables LSTM networks to efficiently process sequential data while maintaining robust long-term memory capabilities, making them particularly effective for tasks requiring the modeling of temporal dependencies.

In this study, we utilize LSTM networks as an alternative to the traditional MCMC method for cosmological parameter estimation. The proposed architecture consists of three hidden LSTM layers followed by a fully connected layer (FCL). Within this framework, the hidden states of the LSTM layers are progressively transformed and projected onto the final output dimensions. Observational data are fed into the network as input, and the information is sequentially processed through each LSTM layer. The output from the final hidden layer is then passed through the FCL to generate estimates of the cosmological parameters, as illustrated in Fig. \ref{fig:nn_model}. This architecture ensures effective feature extraction and accurate parameter inference from the input data.

\subsection{Efficient-Kolmogorov-Arnold networks }\label{2.2}
\begin{figure}
\centering
\includegraphics[width=0.49\textwidth]{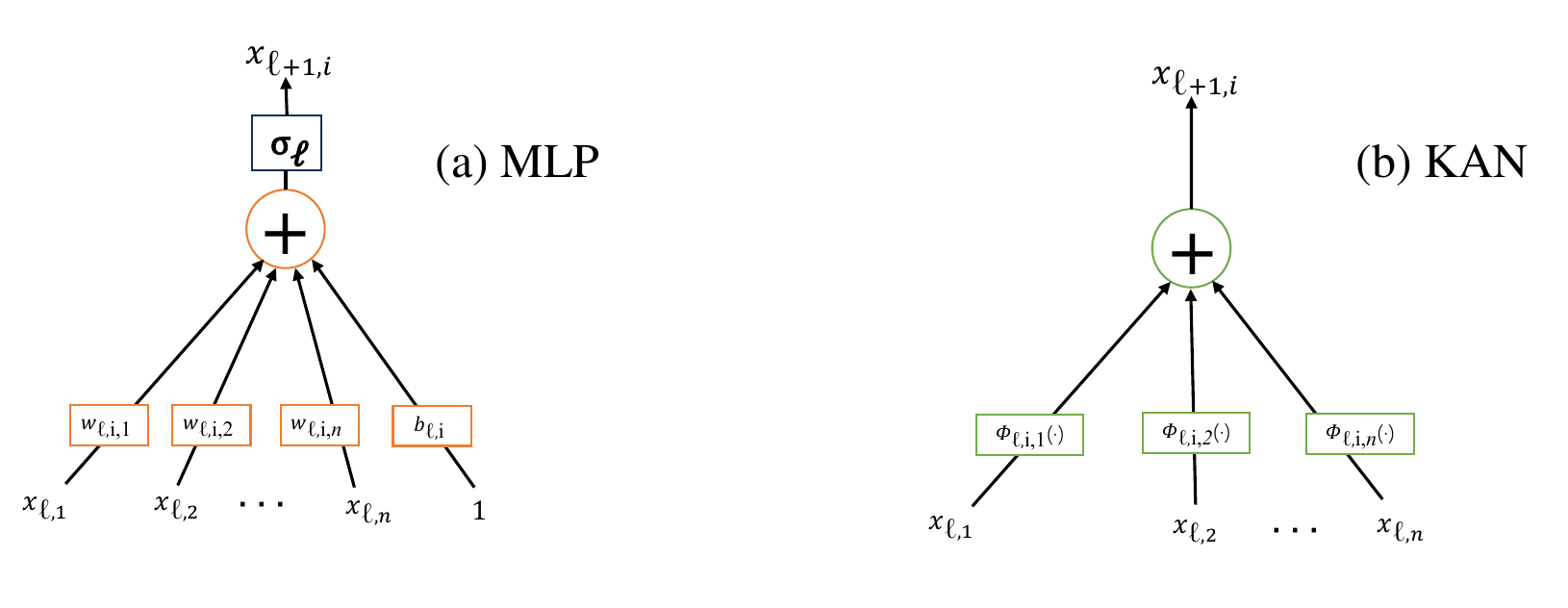}
\caption{ \textit{Left:} Diagram of a single neuron in a Multilayer Perceptron network. \textit{Right:} Diagram of a single neuron in a KAN, where the boxes represent one-dimensional functions.}\label{fig:mlpvskan}
\end{figure}

Kolmogorov-Arnold Networks (KAN) represent a novel alternative to conventional multi-layer perceptrons (MLPs), as proposed by \citet{Liu2024}. This innovative framework integrates B-spline functions with MLPs, offering a unique approach to neural network design. The Kolmogorov-Arnold theorem provides the theoretical foundation for the ideal initialization of KAN, where the ``spatial elements" are equipped with learning-based activation functions, distinct from those used in conventional MLPs. Structurally, KAN adopts a linear (``axis") configuration, with each layer comprising a linear transformation followed by nonlinear activation functions. The group of activation functions can be formally expressed as:
\begin{equation}
X_\ell = \sigma_\ell(W_\ell X_{\ell-1} + b_\ell),
\end{equation}
where $\ell \in \{1, \dots, L\}$ is the layer index, $W_\ell$ is the weight matrix, $b_\ell$ is the bias vector, and $X_\ell$ is the input of the $\ell$-th layer, where $\sigma_\ell$ is the activation function and can choose rectified linear unit, exponential linear unit, logistic sigmoid. In contrast, KAN proposes a different layer operation, where the weight and input are replaced by the trained one-dimensional function $\phi_{\ell,i,j}(X_\ell)$ \citep{Vaswani2017}, as shown in Fig. \ref{fig:mlpvskan} \citep{Liu2024}.
Each item in the output quantity $x_{e+1,i}$ is obtained by calculating:
\begin{equation}
x_{\ell+1,i} = \sum_{j \in \{1, \dots, n\}} \phi_{\ell,i,j}(x_{\ell,j}).
\end{equation}
This can be computed through the matrix-function expression:
\begin{equation}
X_{\ell+1} = \Phi_\ell (X_\ell),
\end{equation}
which provides a convenient expression for $X_\ell$. $\Phi_\ell$ includes the functions for each layer, and these functions are applied to each of the elements of $X_\ell$ and can be described by their unique nonlinear transformation methods.

However, the implementation of the traditional KAN model necessitates expanding the input into large tensors and computing diverse activation functions, resulting in substantial memory consumption. To mitigate this issue, we employ an optimized version termed \textit{Efficient-KAN}. Ef-KAN utilizes an enhanced nonlinear activation function, which reduces computational complexity and optimizes gradient propagation, thereby facilitating more efficient network training. Furthermore, Ef-KAN incorporates a novel regularization technique—weight-based L1 regularization—to enhance generalization capabilities while preserving KAN's robust performance in high-dimensional data modeling. Although Ef-KAN retains the core network architecture of KAN, its optimization strategies enable more efficient computations and more stable training processes.

\begin{figure}
\centering
\includegraphics[width=8.5cm,height=6.5cm]{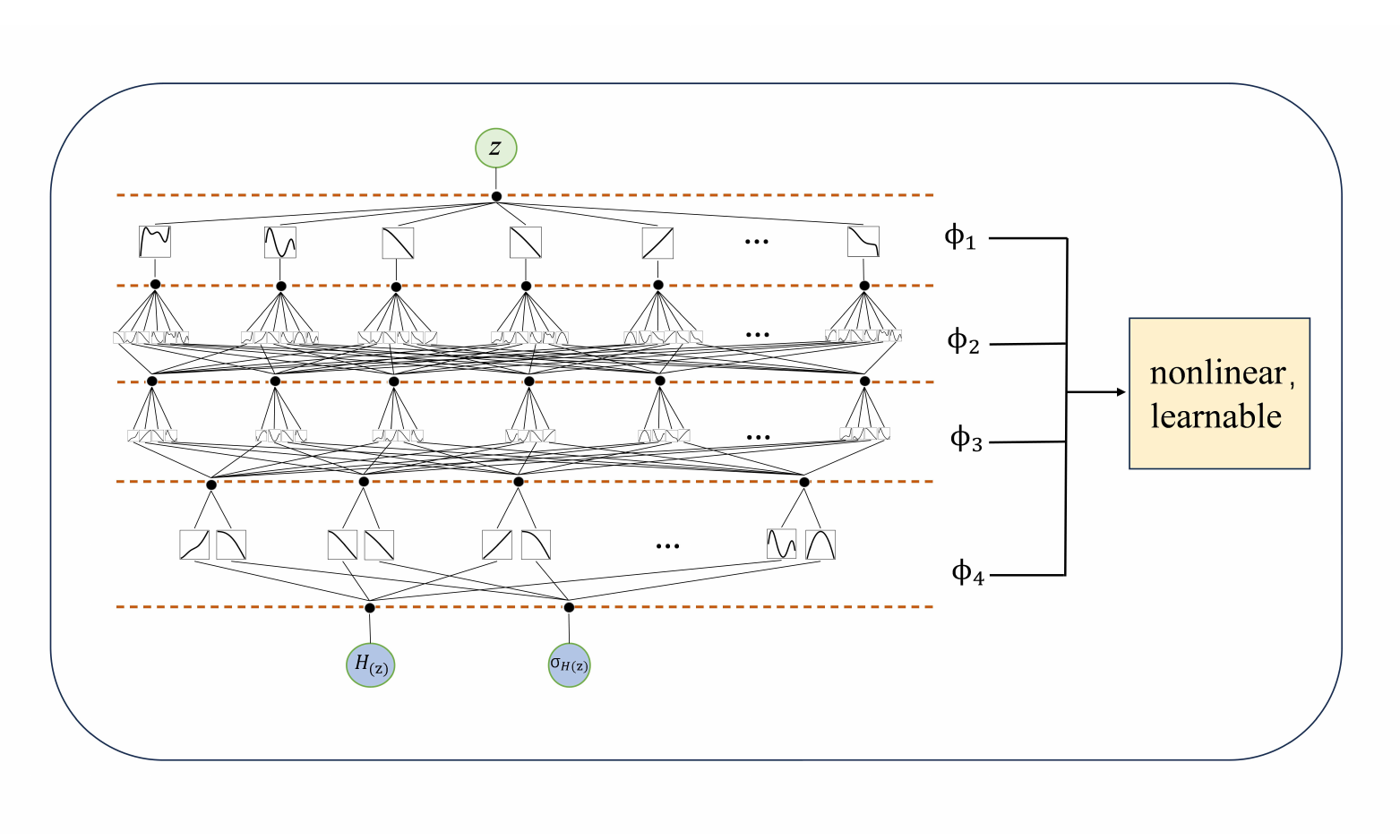}
\caption{The structure of the Ef-KAN, which includes input layers and multiple hidden layers ($\Phi_1$, $\Phi_2$, $\Phi_3$,$\Phi_4$) and output layers. In the hidden layers, the connections between the nodes are represented by learned spline functions, which allow the network to better capture the complex relationships in the data.}\label{fig:figure3}
\end{figure}

In the neural network architecture developed for this study, we utilize four hidden layers of the Ef-KAN structure, as illustrated in Fig. \ref{fig:figure3}. Ef-KAN builds upon the traditional continuous activation function framework but represents each cosmological parameter as a spline-based activation function. This approach allows the network to learn splines as activation functions, enabling the modeling of more complex and nonlinear relationships. Ef-KAN applies splines to the hidden layers and, through specialized training, effectively captures intricate patterns within the data. As demonstrated in our research, Ef-KAN provides a more accurate representation of cosmological structures by addressing challenges related to data fitting and structural complexity, thereby achieving higher precision in parameter estimation and improved interpretability. Additionally, the Ef-KAN architecture offers a clearer and more intuitive design compared to traditional neural networks, enhancing both usability and understanding.

\section{DATA}\label{data}
In this section, we introduce the real OHD and the method for mock $H(z)$ data.
\subsection{The real OHD}\label{ohd}
The measurement of the Hubble parameter $H(z)$ is essential for  studying the evolution and energy density of the universe, as it offers valuable insights into the core properties of the Universe's expansion rate. The primary method for obtaining $H(z)$ is through various observational techniques.
\begin{table}
\setlength{\tabcolsep}{1mm}
\centering \caption{Measurements of $H(z)$ data using the age difference between different galaxy populations.}\label{tab:Hz}
\begin{tabular}{ccc}
\hline
\hline
z 		& $H(z)$ (km $\rm s^{-1}$ $\rm Mpc^{-1}$) & References \\
            \hline
            0.09 	&	$69\pm12$		&   \citet{Jimenez:2003} \\
            \hline
            0.17	&	$83\pm8$		&	\\
            0.27	&	$77\pm14$		&	\\
            0.4		&	$95\pm17$		&	\\
            0.9		&	$117\pm23$		&   \citet{Simon:2005} \\
            1.3		&	$168\pm17$		&	\\
            1.43	&	$177\pm18$		&	\\
            1.53	&	$140\pm14$		&	\\
            1.75	&	$202\pm40$		&	\\
            \hline
            0.48	&	$97\pm62$		&   \citet{Stern:2010} \\
            0.88	&	$90\pm40$		&	\\
            \hline
            0.1791	&	$75\pm4$		&	\\
            0.1993	&	$75\pm5$		&	\\
            0.3519	&	$83\pm14$		&	\\
            0.5929	&	$104\pm13$		&	\citet{Moresco:2012} \\
            0.6797	&	$92\pm8$		&	\\
            0.7812	&	$105\pm12$		&	\\
            0.8754	&	$125\pm17$		&	\\
            1.037	&	$154\pm20$		&	\\
            \hline
            0.07	&	$69\pm19.6$		&	\\
            0.12	&	$68.6\pm26.2$	&	\citet{Zhang:2014} \\
            0.2		&	$72.9\pm29.6$	&	\\
            0.28	&	$88.8\pm36.6$	&	\\
            \hline
            1.363	&	$160\pm33.6$	&	\citet{Moresco:2015} \\
            1.965	&	$186.5\pm50.4$	&	\\
            \hline
            0.3802	&	$83\pm13.5$		&	\\
            0.4004	&	$77\pm10.2$		&	\\
            0.4247	&	$87.1\pm11.2$	&	\citet{Moresco:2016} \\
            0.44497	&	$92.8\pm12.9$	&	\\
            0.4783	&	$80.9\pm9$		&	\\
            \hline
            0.47    &   $89\pm49.6$     &  \citet{Ratsimbazafy:2017} \\
            \hline
            0.80    &   $113.1\pm28.2$  &   \citet{Jiao2023}\\
              \hline
\end{tabular}
\end{table}

One approach to estimating the Hubble parameter  $H(z)$ involves measuring the Baryon Acoustic Oscillations (BAOs) feature \citep{Gaztanaga2009,Blake:2012,Samushia2013}. However, the  $H(z)$ data obtained from the BAO method are inherently contingent upon specific cosmological model assumptions, particularly requiring precise knowledge of the comoving BAO scale, such as the sound horizon. This reliance on model-dependent parameters may limit the applicability of BAO-based  $H(z)$ measurements in model-independent cosmological analyses.

An alternative method employs cosmic chronometers, which estimate the evolution of  $H(z)$ by analyzing the age discrepancy between galaxy populations evolving passively at varying redshifts. This approach offers a significant advantage, as it does not rely on cosmological model assumptions \citep{Jimenez:2002,Jesus:2018}. Specifically, the Hubble parameter is derived from the differential redshift evolution $\Delta z / \Delta t$ of these galaxies, providing a direct and model-independent estimate of  $H(z)$. This makes the cosmic chronometer method particularly valuable for testing cosmological models and constraining the expansion history of the Universe. The Hubble parameter is calculated as:
\begin{equation}
H(z) \approx - \frac{1}{1 + z} \frac{\Delta z}{\Delta t}.
\end{equation}
where $\Delta z$ represents the redshift difference and $\Delta t$ corresponds to the age difference between the galaxy populations. This formula makes the determination of the Hubble parameters reliable and does not require any cosmological assumptions.

We have utilized the most recent compilation of 32 $H(z)$ measurements \citep{Jimenez:2003,Moresco:2012,Moresco:2015,Moresco:2016}, including the latest  $H(z)$ measurements provided by \citet{Jiao2023} based on LEGAC-C survey observations, through an analysis of the spectra of 350 luminous red galaxies.
Based on the above data, the currently available $H(z)$ data points for analysis total 32, as shown in Table \ref{tab:Hz}. This dataset covers redshift range from 0.07 to 1.965. The new measurements by \citet{Jiao2023} at $z = 0.80$ provide more precise data, supporting further research on the evolution of Universe.
\begin{figure}
        \centering
        \includegraphics[width=0.45\textwidth]{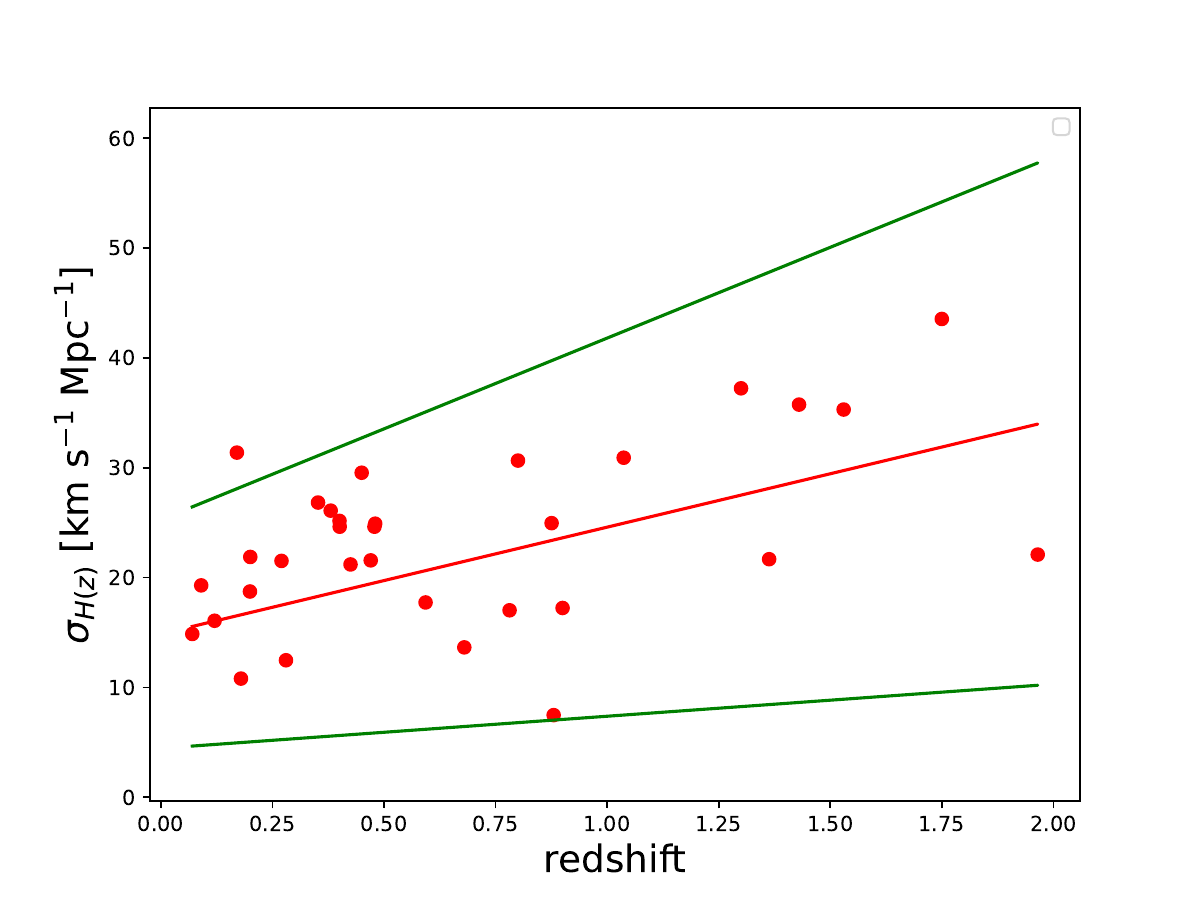}
        \caption{The error of the mock $H(z)$ reconstructed via linear regression. The red solid dots represent the observation error of $H(z)$. The heuristic boundaries $\sigma_+$ and $\sigma_-$ are shown by two green solid lines. The solid line represents the estimated average uncertainty $\sigma_z$.}\label{fig:fenbu}
    \end{figure}
\subsection{The mock $H(z)$ dataset}\label{mockohd}	
We assume a flat $\Lambda$CDM model and adopt the Hubble constant $H_0 = 67.4 \, \text{km s}^{-1} \, \text{Mpc}^{-1}$, and the matter density parameter $\Omega_m = 0.31$ \citep{Aghanim2018} to mock $H(z)$ dataset, which is:
\begin{equation}\label{equ:fLCDM_Hz}
H(z) = H_0 \sqrt{\Omega_{\rm m}(1+z)^3 + \Omega_{\rm\Lambda}}~,
\end{equation}
where cosmological constant term is  $\Omega_{\rm\Lambda}=1-\Omega_{\rm m}$. We assume that the mock $H(z)$ dataset of redshift distribution follows a Gamma distribution, which approximates the observed $H(z)$ dataset in Table \ref{tab:Hz}:
\begin{equation}\label{equ:gammadistribution}
p(x;\alpha,\lambda)=\frac{\lambda^{\alpha}}{\Gamma(\alpha)}x^{\alpha-1}e^{-\lambda x}~,
\end{equation}
where $\alpha$ and $\lambda$ are shape parameter and scale parameter, respectively, and  $\Gamma$ is the Gamma function:
\begin{equation}
\Gamma (\alpha) = \int_{0}^{\infty} e^{-x}x^{\alpha-1}dx~.
\end{equation}
Following the method by \citet{Wang2020ecopann}, we  assume a linear relationship between the error in $H(z)$ and redshift. Specifically, a polynomial fit yields $\sigma_0 = 9.72z + 14.87$ (blue dashed line), representing the average error of $H(z)$ at a given redshift. Next, we select two symmetric green solid lines: $\sigma_- = 2.92z + 4.46$ and $\sigma_+ = 16.52z + 25.28$, ensuring that most data points lie within this region. Finally, we generate random errors $\tilde{\sigma}(z)$ using a Gaussian distribution $N(\sigma_0(z), \epsilon^2(z))$, where $\epsilon(z) = (\sigma_+ - \sigma_-)/4$, ensuring that $\tilde{\sigma}(z)$ falls within this region with 95\% probability.
By employing Eq. \ref{equ:fLCDM_Hz}, we generate the standard value of the Hubble parameter $H_{\text{fid}}(z)$. This standard value is adjusted by adding a random simulation $\Delta H$ that follows a normal distribution $N(0, \tilde{\sigma}(z))$. Therefore, the final simulated Hubble parameter can be generated using the formula:
\begin{equation}
H_{\text{mock},i} = H_{\text{fid}}(z_i) + \Delta H_i,
\end{equation}
The uncertainty is represented by the unknown quantity $\tilde{\sigma}(z)$. This setup enables the generation of simulated Hubble parameter samples with expected redshift error distributions based on the assumed $\Lambda$CDM model. The final mock 32 $H(z)$ data points and their corresponding uncertainties are illustrated in Fig. \ref{fig:fenbu}.

\section{LSTM parameter inference process}\label{LSTMINFERENCE}
In this section, we first provide a brief introduction to the commonly used traditional parameter inference method, MCMC. Then, we will give a detailed introduction to our LSTM parameter inference program.
\subsection{Traditional MCMC parameter inference}
To better illustrate traditional parameter inference methods, we use the OHD as a case study. The observational dataset is defined as $\mathbf{z} = (z_1, \cdots, z_N)^T$, with the corresponding Hubble parameter $H(z)$ represented as $\mathbf{H_{\text{obs}}} = (H_{\text{obs}, 1}, \cdots, H_{\text{obs}, N})^T$, where $N$ is the number of observed $H(z)$ data points. The free parameters  to be derived from $\mathbf{H_{\text{obs}}}$ and $\mathbf{z}$ are represented as a $D$-dimensional vector $\boldsymbol{\theta} = (\theta_1, \cdots, \theta_D)^T$. For instance, for the flat $\Lambda$CDM model, $\boldsymbol{\theta} = (H_0, \Omega_m)^T$. To determine the values of these parameters, it is necessary to estimate the likelihood $P(\mathbf{H_{\text{obs}}} | \boldsymbol{\theta})$. Given the prior $P(\boldsymbol{\theta})$, the posterior distribution of the parameters $P(\boldsymbol{\theta} | \mathbf{H_{\text{obs}}})$ can be determined. For OHD, it is commonly assumed that the data follow a Gaussian distribution, and thus parameter inference is traditionally performed through $\chi^2$ analysis. Assuming the data errors $\mathbf{H_{\text{obs}}}$ have a covariance matrix $\mathbf{\sigma} = (\sigma_i)^T$ , which conforms to a Gaussian distribution with a diagonal covariance matrix (as assumed in e.g., \citep{Ma2011}), the likelihood $L(\boldsymbol{\theta})$ can be written as:
\begin{equation}\label{siran}
\mathcal{L}(\boldsymbol{\theta}) = P(\mathbf{H_{\text{obs}}} | \boldsymbol{\theta}) = \left( \prod_i \frac{1}{\sqrt{2\pi \sigma_i^2}} \right) \exp \left( -\frac{\chi^2}{2} \right),
 \end{equation}
where the $\chi^2$ statistic is:
\begin{equation}
\chi^2 = \sum_i \frac{\left[ H_{\text{th}}(z_i; \boldsymbol{\theta}) - H_{\text{obs},i} \right]^2}{\sigma_i^2}.
\end{equation}
Furthermore, $H_{\text{th}}(z_i; \boldsymbol{\theta})$ is the Hubble parameter at $z_i$ predicted by the theoretical model.  Under the prior knowledge distribution $P(\boldsymbol{\theta})$, the posterior distribution is obtained through Bayesian inference:
\begin{equation}
P(\boldsymbol{\theta} | H_{\text{obs}}) \propto P(H_{\text{obs}} |\boldsymbol{\theta}) P(\boldsymbol{\theta}).
\end{equation}
Based on the above formula, we evaluate the likelihood and calculate the posterior distribution. Given that the parameter space is typically high-dimensional, this process necessitates multidimensional integration. Consequently, the likelihood $\mathcal{L}(\theta)$ is commonly estimated using MCMC methods \citep{Lewis2002,Christensen2001}. MCMC constructs a stochastic random walk through the parameter space, approximating the posterior distribution (e.g., a Gaussian distribution). The procedure involves initializing the parameters, defining the target distribution (analogous to the prior distribution), and employing sampling algorithms (such as the Metropolis-Hastings method) to guide the chain toward high-probability regions of the parameter space, thereby generating samples from the posterior distribution. The chain eventually converges to the most probable parameter values, with statistical metrics such as the $\chi^2$ method used to estimate parameter uncertainties. However, MCMC methods, which rely on$\chi^2$ analysis, encounter challenges when processing complex likelihood functions, particularly those that are computationally intractable or highly non-linear. To address these limitations, recent studies have proposed an LSTM-based neural network model for cosmological parameter inference, offering a more efficient and scalable alternative to traditional MCMC approaches.

\subsection{LSTM network and hyperparameters}\label{Hyperparameters}
In addressing parameter estimation tasks, we utilized a neural network architecture based on LSTM network to capture the intricate relationships between observational data and cosmological parameters, with the final output mapping achieved through fully connected layers. Prior to conducting parameter estimation, it is essential to select an appropriate set of hyperparameters, including the number of input and output nodes, the dimensionality of hidden states in the LSTM layers, the number of network layers, the learning rate, and the batch size. Currently, there is no definitive experimental evidence or theoretical framework to determine the optimal network structure for a specific task; thus, the design of the neural network architecture often relies on empirical knowledge and iterative experimentation.

For this study, we constructed a stacked network consisting of three LSTM layers, each configured with a hidden state dimension of 256. Between each LSTM layer, we incorporated dropout operations with a rate of 0.25. This means that during each forward pass in the training process, 25\% of the neurons in each LSTM layer are randomly deactivated (set to zero). This technique reduces the model's over-reliance on specific neurons, mitigates the risk of overfitting, and enhances the model's generalization capabilities on unseen data. The network's output is derived from the LSTM features at the last time step and is subsequently fed into a fully connected layer. This fully connected layer maps the high-dimensional features into the target output space, enabling precise parameter estimation.

The learning rate, a critical hyperparameter for adjusting the network's weights, was initially set to $10^{-3}$ in our experiments. During the training process, it is gradually decayed to $10^{-8}$. This approach ensures rapid convergence during the initial stages of training while allowing the model to fine-tune and approach the optimal solution more precisely in later stages. The batch size is dynamically adjusted based on the number of training samples to maintain a fixed number of iterations per training epoch. The training process is divided into several rounds, with each round consisting of a consistent number of iterations, thereby ensuring a sufficient total number of iterations for robust model training. This comprehensive training strategy stabilizes the loss function and effectively promotes model convergence, ultimately enhancing the accuracy and reliability of the parameter estimation.

\subsection{Training set}\label{Training Set}
LSTM networks are used in this study to learn the relationship between observed data and cosmological parameters. To ensure that LSTM can make accurate predictions for observational data, it is trained to predict all possible true values of the parameters. In this study, we follow the method of \citet{Wang2020ecopann}, where the parameter values are taken from the range $[P - 5\sigma_p, P + 5\sigma_p]$, with $P$ representing the mean of the posterior distribution of the parameter and $\sigma_p$ being the corresponding 1$\sigma$ error. This range of parameters space is designed to effectively cover the posterior distribution of the parameter. During training, we ensure that the LSTM model generalizes well across this parameter space, thereby enhancing the training performance of the LSTM.

\subsection{Data preprocessing and noise enhancement strategy}\label{4.4}
In supervised learning, in order to improve the model's ability to generalize to test data, the training data needs to match the distribution of the test data. In this study, the test set represents the observed data. These data typically exhibit measurement errors, which arise from measurement discrepancies and are modeled as $N(\overline{X}, \sigma^2)$, where $\overline{X}$ is the mean of the observed data, and $\sigma$ is the corresponding measurement error. Therefore, to enhance the model's generalization capability, we perform preprocessing and augmentation to ensure a better match of the observed data distribution.

Due to the differences in the magnitudes of different cosmological parameters, which may affect the training results, we first perform parameter scaling on the training data to reduce the feature values, bringing their range closer to 1. Then, we apply the Z-Score normalization method for further processing of the training data.
\begin{equation}\label{equ:24}
z = \frac{x - \mu}{\sigma},
\end{equation}
where $\mu$ and $\sigma$ represent the mean and standard deviation of the observed value $X$ or the cosmological parameter $P$.

To avoid the discrepancies between the training data and the true observational data distribution, which may reduce the model's performance, we add high-frequency noise during the training phase \citep{Wang2020ecopann}. According to research \citep{Bishop1995}, noise is introduced into the data as a form of regularization, which helps improve the model's generalization ability. In each training iteration, Gaussian noise $N(0, A\sigma^2)$ is added to each sample, where $A$ is a constant. The noise samples are different in each iteration, and after a sufficient number of training iterations, the LSTM model can better learn the characteristics of the observational data. Furthermore, to further reduce small network-specific experimental data deviations, we assume $A$ follows a distribution $N(0, 0.25\sigma^2)$ \citep{Wang2020ecopann}.
\subsection{Training process and parameter inference}\label{4.5}
After the preprocessing operation in the previous section, we  outline our method for training on the  LSTM parameter inference model and predict cosmological parameters.

1. \textit{Initializing the parameter space:} Use the specified initial range as the initial parameter space for cosmological parameters;

2. \textit{Simulating training data:} In this process, we use the method described in Section \ref{Training Set} to generate a large number of parameter samples in the parameter space, through which the model obtains simulated training data corresponding to the theoretical model;

3. \textit{Training the LSTM parameter inference model:} The model generated in step 2 is processed through the method described in Section \ref{4.4}, and these data are used to train the LSTM neural network. After training, the neural network can predict the parameter values from the simulated data;

4. \textit{Parameter prediction:} Use the trained LSTM parameter inference model to predict cosmological parameters for actual observational data through multiple iterations of predictions:  Firstly, the model makes predictions in a sequential manner, generating a variety of different predictions; Then, each group of observations produces corresponding LSTM model outputs, yielding the predicted set of parameters.

We generate parameter prediction results and construct a parameter sample chain, i.e., comprising approximately 10,000 samples of $(H_0, \Omega_m)$. It is crucial to note that the first sample in the chain must be generated based on the parameters obtained after training the initial parameter space model, as the training data may exhibit biases in the parameter sample distribution. Although the initial parameter samples in the chain may not fully align with the true distribution of the observational data, they are still utilized to approximate the true posterior distribution. For instance, in the flat $\Lambda$CDM model, if the distribution of the parameter $H_0$ is centered around the prior and the parameter range does not require extensive exploration, the range of $H_0$ can be estimated as $[70 - \Delta, 70 + \Delta]$, where $\Delta$ is a specific constant. A similar procedure is applied to other parameters. After an additional training iteration, the subsequent model data will more closely approximate the true distribution of cosmological parameters, with the sample chain converging toward the true posterior distribution in the parameter space. This iterative refinement of the parameter space allows us to augment the training data, thereby enhancing the precision of the LSTM model's training.

Through repeated iterations of this process, the parameter space gradually converges. When a predefined convergence criterion is met (the variation in parameter estimates becomes sufficiently small), we combine all generated parameter chains for subsequent parameter inference tasks. This approach ensures robust and accurate parameter estimation by iteratively refining the parameter space and improving the model's alignment with the true underlying distribution.

\subsection{Applications for OHD}
\begin{figure}
\begin{center}
\includegraphics[width=0.46\textwidth]{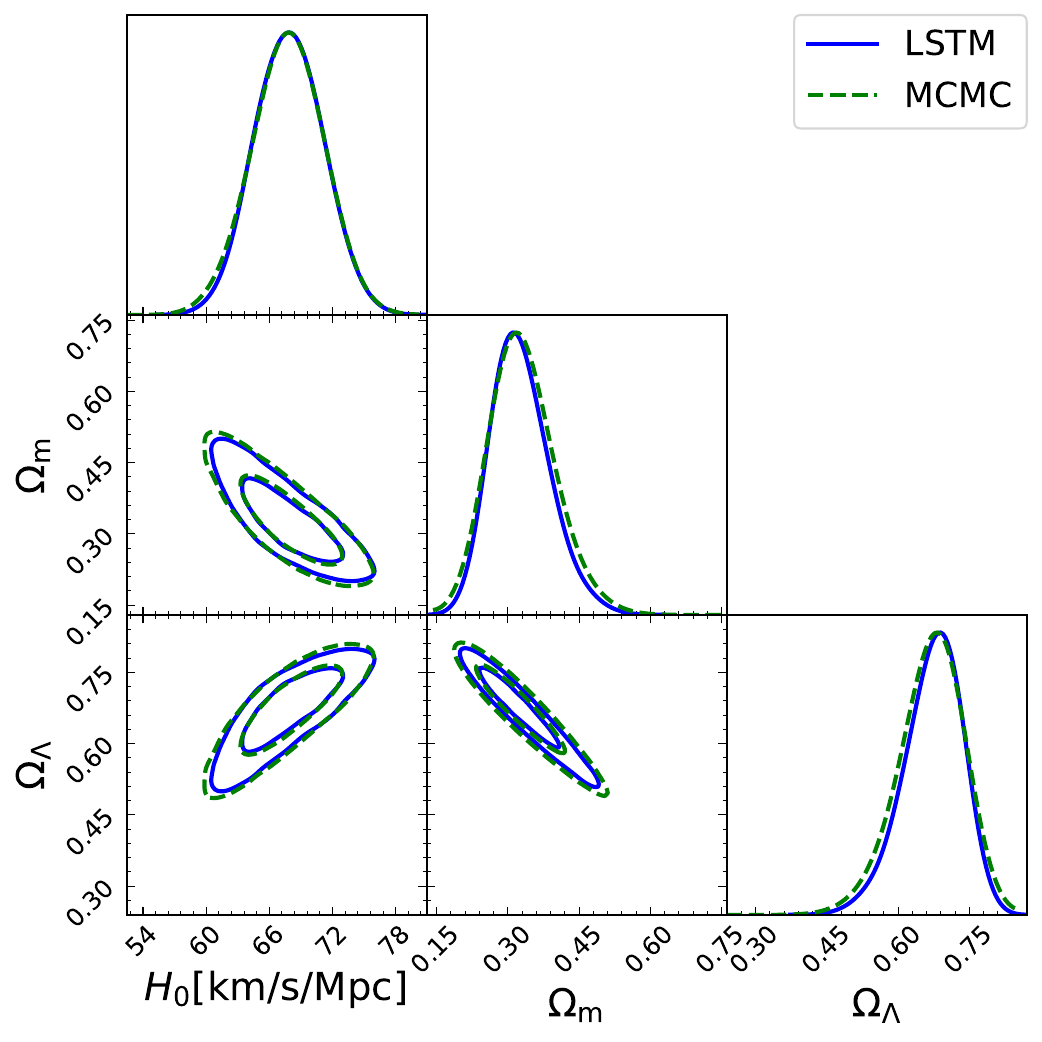}
\end{center}
\caption{The 2-D plots and 1-D marginalized distributions with 1-$\sigma$ and 2-$\sigma$ contours of cosmological parameters using the LSTM method (blue solid line) and MCMC method (green dashed line) from 32 real OHD.}\label{fig:lstm-mcmc}
\end{figure}
In our analysis, the model parameters used to infer the Hubble parameter rely on the method described in Section \ref{mockohd}. In the $\Lambda$CDM cosmological model, the fiducial values of the cosmological parameters are set as $H_0 = 67.4 \, \text{km s}^{-1} \, \text{Mpc}^{-1}$ and $\Omega_m = 0.314$. In this case, only two key parameters, namely the Hubble constant $H_0$ and the matter density parameter $\Omega_m$, are required for the analysis. First, we use the Python package emcee \citep{ForemanMackey2013} to perform the analysis of the cosmological parameters using the MCMC method. In this process, we generate 100,000 MCMC chains and use the corner tool to analyze these chains and compute the best-fit values for each parameter, along with their corresponding 1$\sigma$ uncertainties (see Table \ref{tab:lstm}).

Secondly, we employ the LSTM method (for a detailed description, see Section \ref{2.1}) to independently fit the cosmological parameters. Specifically, we use the OHD as input to predict the values of two cosmological parameters. In practice, to determine the initial ranges of these two parameters, we set the range for $H_0$ to $[0, 100]$, and the initial range for $\Omega_m$ to $[0.0, 0.595]$. After training, the LSTM model takes OHD as input and outputs the predicted values of $H_0$ and $\Omega_m$.

During the model training process, we used 2,000 simulated $H(z)$ data sets to initially train the LSTM, and following the specific training procedure outlined in Section \ref{4.5}, we ultimately obtained 3 cosmological parameter chains. Based on these results, we calculated the best-fit values of the parameters along with their 1$\sigma$ error ranges and compared and analyzed these with the results from the MCMC method. The results show that the overall deviations between the LSTM method and the MCMC method are very close, with the relative deviation for $H_0$ being only 0.128\%, and for $\Omega_m$ it is 2.51\%. This indicates that the results from the LSTM method are highly consistent with those from the MCMC method. We also separately calculated the deviations of the parameter inference results from the MCMC and LSTM models compared to the benchmark. Specifically for $H_0$, the absolute deviation of the MCMC method is 0.148$\sigma$, while the LSTM method has a slightly smaller absolute deviation of 0.118$\sigma$, bringing it closer to the benchmark value. For $\Omega_m$, the absolute deviation of the MCMC method is 0.045$\sigma$, and for the LSTM method, it is 0.071$\sigma$. Additionally, to visually present the results, we plotted Fig. \ref{fig:lstm-mcmc}, where the fitting results from the LSTM method almost completely coincide with those from the MCMC method. Thus, the LSTM method performs well in constraining cosmological parameters, effectively reproducing the analysis results from the MCMC method, thereby further demonstrating the reliability of the experimental results and the credibility of the method.
\begin{table}
\setlength{\tabcolsep}{1.5pt}
\setlength{\tabcolsep}{5mm}
 \caption{\label{tab:lstm} The best-fitting values and 1$\sigma$ uncertainty results of $H_0$, $\Omega_m$ and $\Omega_\Lambda$ using the OHD with the MCMC and LSTM methods.}
\begin{center}
\renewcommand{\arraystretch}{2}
\begin{tabular}{c| c| c }
\hline
\hline
{Parameters} & {MCMC} & {LSTM}
\\
\hline
$H_0$ [km/s/Mpc]  & $67.86^{+3.10}_{-3.22}$ & $67.77^{+3.12}_{-3.02}$ \\
\hline
 $\Omega_m$ & $0.31^{+0.07}_{-0.04}$ & $0.32^{+0.07}_{-0.04}$  \\
\hline
$\Omega_\Lambda$ & $0.69^{+0.04}_{-0.07}$ & $0.68^{+0.04}_{-0.07}$\\
\hline
\hline
\end{tabular}
\end{center}
\end{table}

\section{Reconstruction of $H(z)$ and parameter constraints}\label{sec:reconstruct_Hz}

In this section, we will illustrate how to use the simulated data generated in Section \ref{mockohd} to train and predict Hubble parameters in the redshift range of [0, 2] with the Ef-KAN model.

\subsection{Ef-KAN model training}\label{xunlian}
\begin{figure*}
\centering
\includegraphics[width=0.9\textwidth]{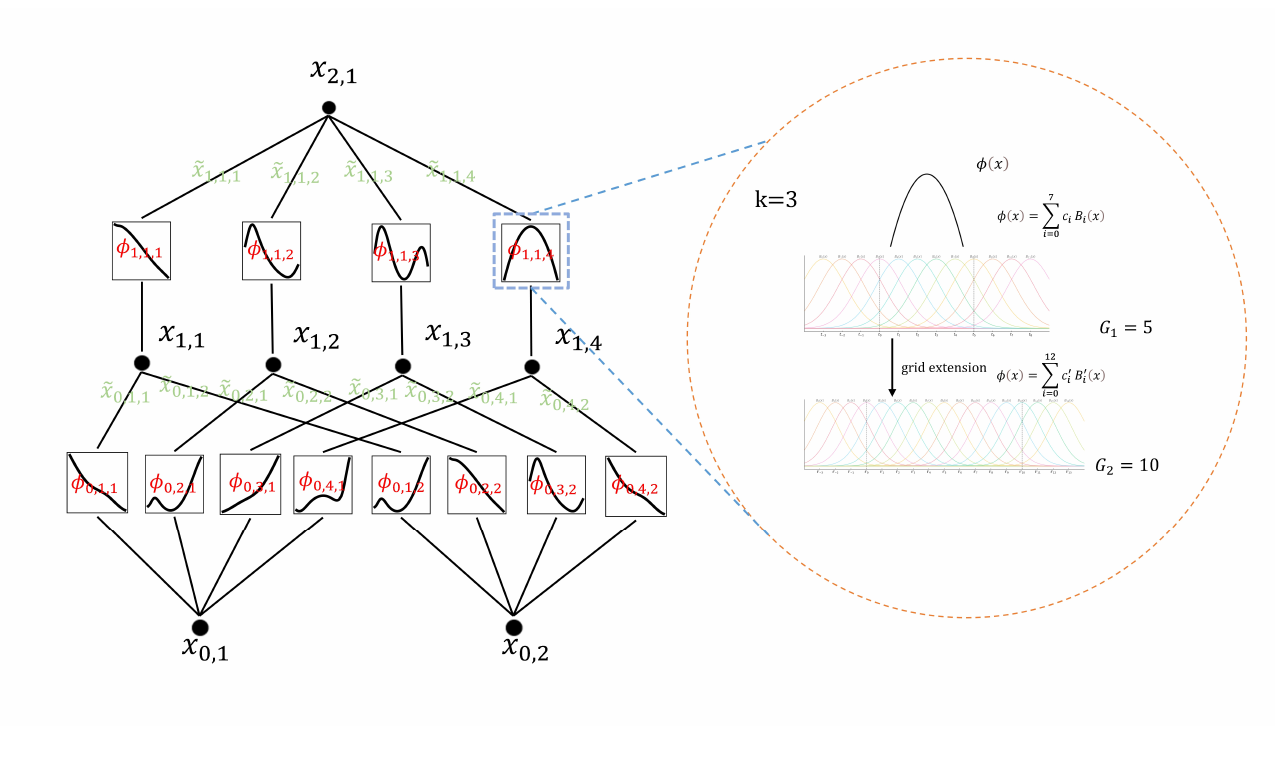}
\caption{\textit{Left}: Symbols of activations flowing through the network. \textit{Right}: The activation function is parameterized with B-splines, which allows switching between coarse and fine grids.}\label{fig:fuhao}
\setlength{\belowcaptionskip}{1pt}
\end{figure*}
The goal of Ef-KAN is to construct an approximate function model based on the mapping relationship between input and output data. In this section, Ef-KAN relates redshift $z$ with the Hubble parameter $H(z)$ and the uncertainty based on $H(z)$ data. Therefore, the input to the Ef-KAN model is redshift $z$, and the outputs are the corresponding $H(z)$ values and their errors $\sigma_{H(z)}$. Before inputting the data into the Ef-KAN model, we performed Z-score normalization on the data according to the Eq. (\ref{equ:24}) given in Section \ref{4.4}, and after output, we performed inverse normalization to return the data to its original scale. As mentioned in Section \ref{2.2}, to transform the Kolmogorov-Arnold representation into a learnable neural network model, we need to parameterize it, that is, parameterize the activation functions as B-splines. We introduce basis functions $b(x)$ (similar to residual connections), and the activation function $\phi(x)$ is the sum of $b(x)$ and spline functions:
\begin{equation}
\phi(x) = w_b b(x) + w_s \, \text{spline}(x),
\end{equation}
where $b(x)$ is defined as follows:
\begin{equation}
b(x) = \text{silu}(x) = \frac{x}{1 + e^{-x}},
\end{equation}
and the spline function \texttt{spline$(x)$} is parameterized as a linear combination of B-splines:
\begin{equation}
\text{spline}(x) = \sum_i c_i B_i(x).
\end{equation}
For B-splines, the function maintains the same continuity within its domain and at the knots, and its polynomial expression can be expressed by the Cox-de Boor recursion formula as follows:
\begin{equation}
B_{i,0}(x) :=
\begin{cases}
1 & \text{if } t_i \leq x < t_{i+1} \\
  0 & \text{otherwise},
\end{cases}
\end{equation}

\begin{equation}
B_{i,k}(x) := \frac{x - t_i}{t_{i+k} - t_i} B_{i,k-1} + \frac{t_{i+k+1} - x}{t_{i+k+1} - t_{i+1}} B_{i+1,k-1},
\end{equation}
where $t_i$ is the knot vector that defines the positions of the segments, and $k$ is the degree of the B-spline. For instance, $k = 1$ is linear, $k = 2$ is quadratic, and $k = 3$ is cubic.

Theoretically, $\omega_b$ and $\omega_s$ are redundant parameters because they can be absorbed into $b(x)$ and $\text{spline}(x)$. However, we retain these factors in this work as they are also trainable, allowing for some degree of modulation of the overall amplitude of the activation function. Therefore, as presented in the right diagram of Fig. \ref{fig:fuhao} \citep{Liu2024}, the parameters that Ef-KAN truly needs to learn are the coefficients $c_i$ in front of the B-spline basis functions here. We initialize the activation function with $\omega_b = 1$ and $\text{spline}(x) \approx 0$, and the weights $\omega_0$ are initialized using the Xavier method, similar to initializing an MLP.

In supervised learning tasks, data is commonly split into training, validation, and test sets. The training set is used for model training, The training set is utilized to train the model, the validation set helps to fine-tune hyperparameters (like learning rate, hidden layers, and neurons), and the test set serves as performance evaluation. However, in this task, we used all the simulated $H(z)$ data for training, aiming to approximate a function, so we did not use separate validation and test sets.

Nevertheless, hyperparameter optimization remains an important step in building an efficient and robust machine learning model. Hyperparameters are parameters that are not automatically learned from data during the training process but are manually set by the user. In this study, we primarily focus on the number of hidden layers and the number of neurons per layer. Following the research approach of \citet{Zhang2024}, we employed grid search method to determine the optimal combination of hyperparameters. We set the number of hidden layers in the Ef-KAN network to $[1, 2, 3, 4]$, and the range of neuron counts to $[100, 200, 300, 400, 500, 600, 700]$. The grid search is an exhaustive method that lists all possible hyperparameter combinations, and \citet{lavalle2004} have demonstrated its effectiveness in hyperparameter tuning. Finally, we identified a model with a three-layer Ef-KAN network, with each layer having $[200, 200, 100]$ neurons respectively.

\subsection{Reconstruction of Hubble parameter and comparing with other models }\label{reconstruct}

\begin{figure}
\includegraphics[width=0.48\textwidth]{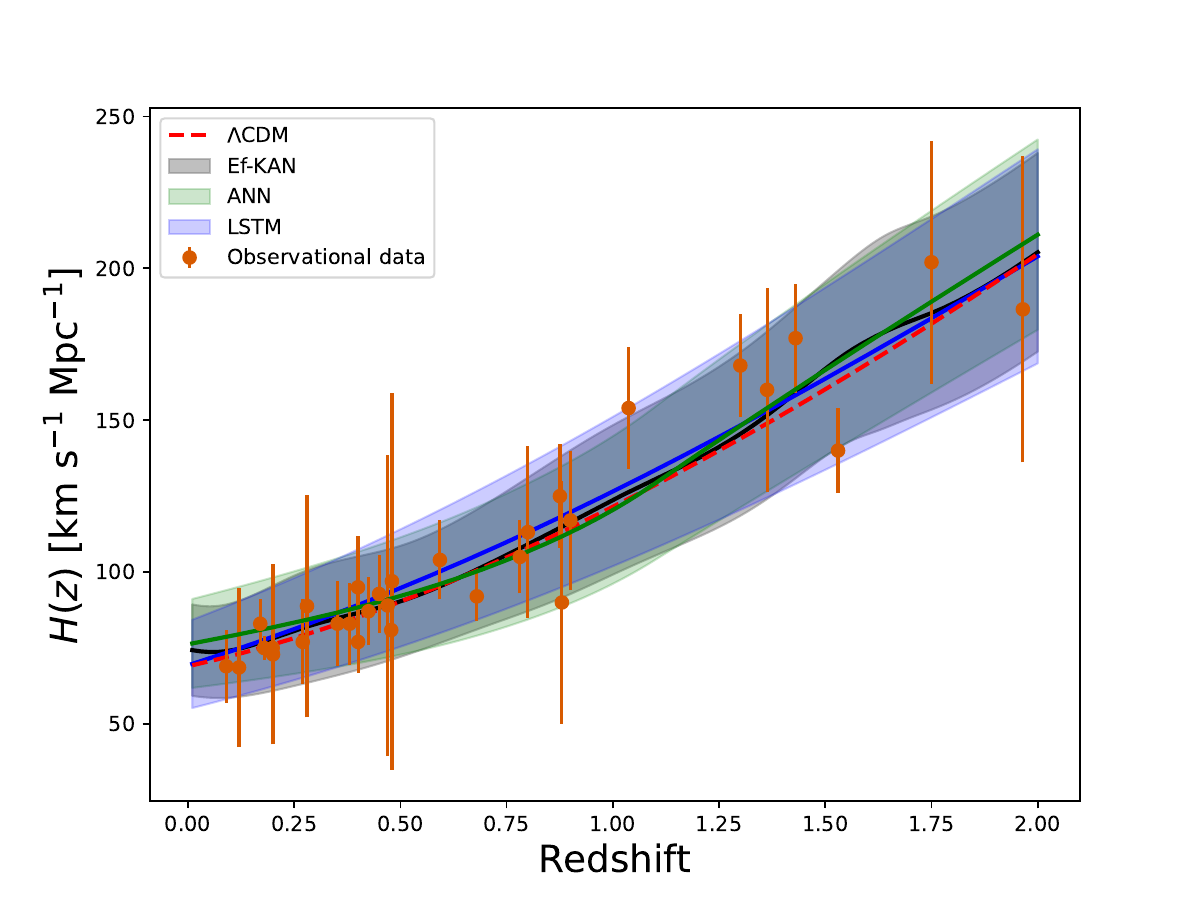}
\caption{The comparison of results for reconstructing the Hubble parameter using different methods.  The gray area and black line represent the $1\sigma$ confidence interval and the best-fitting values for the Ef-KAN model, the green area and green line indicate the corresponding results for the ANN model, the blue area and blue line represent the results for LSTM model, and the red dashed line represents the fiducial $\Lambda$CDM model with the best-fitting values.}\label{fig:vs}
\end{figure}

Using the grid search  method, we obtained the best-performing Ef-KAN model. Using this optimal Ef-KAN model, after training with mock data, we input a series of redshifts into the network model and obtained a sequence of Hubble parameters with errors, effectively reconstructing the Hubble parameters in the  $[0, 2]$ redshift  range. The results are shown in Fig. \ref{fig:vs}.  The reconstruction shows that at low and medium redshifts, the reconstructed Hubble parameters almost coincide with the $\Lambda$CDM standard model, with slight deviations at high redshifts, consistent with the distribution of current actual observational data. Overall, our Ef-KAN model can achieve good reconstruction results when trained with a limited amount of mock  data.

We used the same method described in Section \ref{xunlian} to find the optimal hyperparameters for the network, and identified the optimal models for LSTM and ANN. Similarly, we used mock  Hubble parameters to reconstruct $H(z)$ in the $[0, 2]$ redshift  range, and compared the reconstruction structures with the Ef-KAN model, as shown in Fig. \ref{fig:vs}. Visualization shows that the best-fit values of the Ef-KAN model coincide most with the flat $\Lambda$CDM model. We used coefficient of determination $R^2$ \citep{Fisher1921} to quantify the accuracy of the $H(z)$ reconstruction results. The $R^2$ is a commonly used evaluation metric in regression analysis, measuring the model's explanatory power of the target variable and reflecting the fit between predicted and actual values. The formula for calculating $R^2$ is:
\begin{equation}
R^2 = 1 - \frac{SS_{\text{res}}}{SS_{\text{tot}}},
\end{equation}
where $SS_{\text{tot}}$ is  the total sum of squares and  $SS_{\text{res}}$ is the residual sum of squares.

In general,  the closer the $R^2$ value is to 1, the better the model fits the data, and the more variability of the target variable it can explain; conversely, the closer the $R^2$ value is to 0, the worse the model's explanatory power. Additionally, to more comprehensively evaluate model performance, we also used the Mean Absolute Error (MAE) (\cite{Bickel1977}) and the Akaike Information Criterion (AIC) \citep{Akaike1974}. AIC accounts for both the goodness of fit (likelihood) and the complexity (free parameter count) of the model, where a lower value indicates better model performance. MAE calculates the deviation between the predicted and actual outcomes of a regression model by calculating the average of the absolute errors between predictions and actuals; the smaller it is, the higher the model's predictive accuracy.

From the data in Table \ref{tab:vs}, it is apparent that the Ef-KAN model performs best among all three metrics. Specifically, the Ef-KAN model has an $R^2$ value of 0.9975, indicating excellent fit and a strong ability to explain the variability of the target variable. Additionally, its AIC value is 40.18, significantly lower than the LSTM (57.86) and ANN (81.60) models, indicating that the Ef-KAN model has lower complexity and better goodness of fit. The Ef-KAN model's MAE is 1.43, meaning that the average absolute error between its predicted and actual values is small, further demonstrating its advantage in precision.

In contrast, both the LSTM and ANN models perform worse than the Ef-KAN model across all metrics. Although the LSTM has an $R^2$ value of 0.9962, indicating strong fitting capability, its AIC and MAE values are significantly higher than those of the Ef-KAN model, suggesting that the LSTM lacks in model complexity and predictive accuracy. The ANN model has the lowest $R^2$ value (0.9929) and the highest AIC and MAE values, indicating that it performs the worst among the three models. In summary, the Ef-KAN model performed the best in this experiment, providing the optimal balance between fitting precision, model complexity, and predictive accuracy.

\begin{figure}[htbp]
\centering
\begin{minipage}{0.46\textwidth}
\begin{center}
\includegraphics[width=\textwidth]{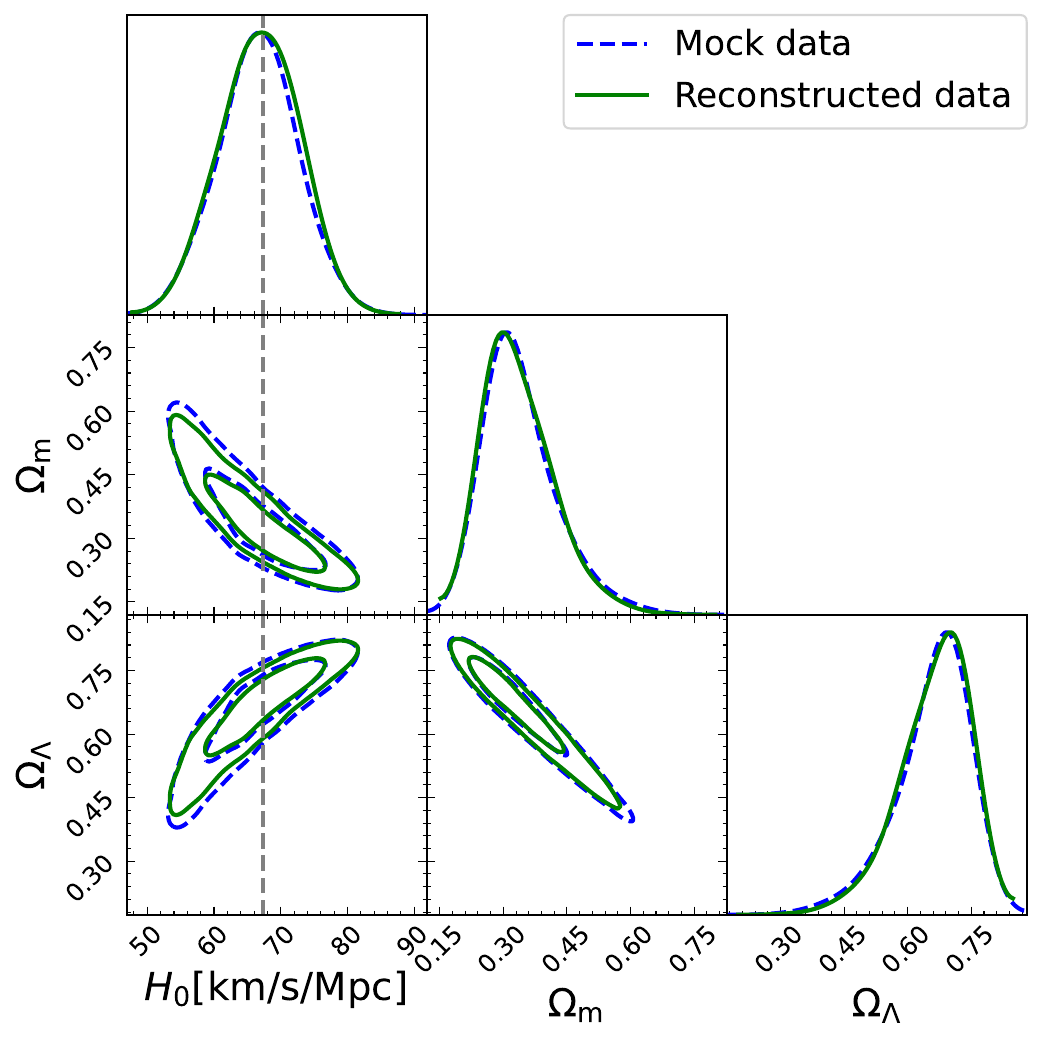}
\end{center}
\end{minipage}
\caption{The LSTM model infers the posterior distribution by randomly selecting 32 sets of data from the reconstructed $H(z)$ data (green solid line) and  from the mock $H(z)$ data (blue dashed line).}
\label{fig:lstm-rec_hz}
\end{figure}

\begin{table}
\setlength{\tabcolsep}{2pt}
\setlength{\tabcolsep}{5mm}
\caption{\label{tab:vs} The performance of the three models (Ef-KAN, LSTM, and ANN) based on $R^{2}$, AIC, and MAE metrics.}
\begin{center}
\renewcommand{\arraystretch}{2}
\begin{tabular}{c| c| c | c }
\hline
\hline
{Model} & $R^2$ & {AIC}& {MAE}
\\
\hline
Ef-KAN & 0.9975 & 40.18 & 1.43 \\
\hline
 LSTM & 0.9962 & 57.86 & 2.23 \\
\hline
ANN & 0.9929 & 81.60 & 3.06 \\
\hline
\hline
\end{tabular}
\end{center}
\end{table}

\subsection{Estimation with reconstructed H(z)}\label{sec:ES_Hz}
Since the reliability of our proposed LSTM parameter inference method was demonstrated in Section \ref{LSTMINFERENCE}, and it outperformed the MCMC method in parameter inference, we employed this LSTM model to constrain cosmological parameters. The model's parameter settings are referenced in Section \ref{Hyperparameters}.
We used 32 mock data to obtain the posterior distribution of cosmological parameters estimated by the LSTM model and the final results are shown in Fig. \ref{fig:lstm-rec_hz}, with results of $H_0 = 66.93^{+5.22}_{-5.88}$ km s$^{-1}$ Mpc$^{-1}$, $\Omega_m = 0.31^{+0.10}_{-0.05}$ and $\Omega_\Lambda = 0.69^{+0.05}_{-0.10}$. To perform a more comprehensive comparison, we also randomly selected 32 data points from the  Ef-KAN reconstructed $H(z)$ data and used the LSTM model to constrain cosmological parameters. The posterior distribution estimated by the LSTM model is also shown in Fig. \ref{fig:lstm-rec_hz}, giving results of $H_0 = 65.54^{+6.57}_{-5.65}$ km s$^{-1}$ Mpc$^{-1}$, $\Omega_m = 0.30^{+0.11}_{-0.04}$ and $\Omega_\Lambda = 0.70^{+0.04}_{-0.11}$.
These results indicate that our proposed LSTM-based parameter inference method is robust and consistent across different datasets, further validating its reliability in cosmological parameter estimation. The close agreement between the results obtained from the mock data and the Ef-KAN reconstructed $H(z)$  data suggests that the LSTM model is capable of accurately capturing the underlying cosmological constraints, even when applied to observational data with inherent uncertainties.

Moreover, the precision of the inferred parameters, particularly the tight constraints on cosmological parameters, highlights the potential of the LSTM model as a powerful tool for cosmological analysis. The slight variations in the inferred values of cosmological parameters between the two datasets are within the expected statistical fluctuations, further reinforcing the consistency of the method.
In addition to the parameter constraints, the posterior distributions shown in Fig. \ref{fig:lstm-rec_hz} provide valuable insights into the degeneracies and correlations between the cosmological parameters. The LSTM model effectively captures these relationships.

\section{Conclusions and Discussions}\label{sec:discussion}
This study demonstrates the effectiveness of LSTM networks and Ef-KAN models in cosmological parameter estimation and Hubble parameter reconstruction. LSTM networks, with their ability to capture long-term dependencies in sequential data, provide a powerful tool for likelihood-free inference, achieving results comparable to traditional MCMC methods. The Ef-KAN model, on the other hand, excels in reconstructing the Hubble parameter  from real observational and mock  data, offering a data-driven approach that does not rely on prior assumptions. Our experiments show that the reconstructed $H(z)$ closely matches the $\Lambda$CDM model, particularly at low and medium redshifts, with slight deviations at higher redshifts, consistent with current observational data.

The LSTM-based parameter inference method outperforms traditional MCMC in terms of computational efficiency and scalability, making it particularly suitable for high-dimensional parameter spaces. The Ef-KAN model, with its ability to model complex, non-linear data distributions, provides a more accurate and interpretable framework for Hubble parameter reconstruction. Together, these methods offer a robust and efficient approach to cosmological data analysis, addressing the limitations of traditional techniques such as GP and ANN.
However, some challenges remain. The training process for LSTM and Ef-KAN networks can be complex, especially when handling high-dimensional data, and further optimization strategies are needed to enhance model stability and generalization. Additionally, while LSTM networks can approximate the true posterior distribution effectively, estimation errors may still occur under extreme conditions. Future studies should aim to combine additional observational data with theoretical models to enhance the effectiveness of these approaches.
Future work could involve applying the LSTM model to larger and more diverse datasets, including those from upcoming cosmological surveys, to further test its robustness and scalability. Additionally, exploring the integration of the LSTM model with other machine learning techniques or traditional statistical methods could enhance its performance and provide even more precise constraints on cosmological parameters.

In conclusion, LSTM networks and Ef-KAN models represent significant advancements in cosmological parameter inference and Hubble parameter reconstruction. As computational capabilities improve and more data becomes available, these data-driven methods are expected to play an increasingly important role in advancing our understanding of the universe. The combination of LSTM and Ef-KAN offers a promising direction for future cosmological research, particularly in addressing the challenges posed by complex datasets and high-dimensional parameter spaces.

\section*{Acknowledgments}
This work was supported by the National Natural Science Foundation of China under Grants No. 12374408, No. 12203009, No. 12475051, and No. 12421005; The science and technology innovation Program of Hunan Province under grant No. 2024RC1050; The Natural Science Foundation of Hunan Province under grant No. 2023JJ30384; and the innovative research group of Hunan Province under Grant No. 2024JJ1006; The Hubei Province Foreign Expert Project (2023DJC040);
\bibliography{reference}{}

\begin{thebibliography}{}
\expandafter\ifx\csname natexlab\endcsname\relax\def\natexlab#1{#1}\fi
\providecommand{\url}[1]{\href{#1}{#1}}
\providecommand{\dodoi}[1]{doi:~\href{http://doi.org/#1}{\nolinkurl{#1}}}
\providecommand{\doeprint}[1]{\href{http://ascl.net/#1}{\nolinkurl{http://ascl.net/#1}}}
\providecommand{\doarXiv}[1]{\href{https://arxiv.org/abs/#1}{\nolinkurl{https://arxiv.org/abs/#1}}}

\bibitem[{Akaike(1974)}]{Akaike1974}
Akaike, H. 1974, IEEE Transactions on Automatic Control, 19, 716,
  \dodoi{10.1109/TAC.1974.1100705}

\bibitem[{{Alsing} {et~al.}(2019){Alsing}, {Charnock}, {Feeney}, \&
  {Wandelt}}]{Alsing2019}
{Alsing}, J., {Charnock}, T., {Feeney}, S., \& {Wandelt}, B. 2019, \mnras, 488,
  4440, \dodoi{10.1093/mnras/stz1960}

\bibitem[{Bickel \& Doksum(1977)}]{Bickel1977}
Bickel, P.~J., \& Doksum, K.~A. 1977, Mathematical Statistics: Basic Ideas and
  Selected Topics, Volume 1 (Holden-Day), 47.
\newblock \url{https://archive.org/details/isbn_0816207844}

\bibitem[{Bishop(1994)}]{Bishop1994}
Bishop, C.~M. 1994

\bibitem[{Bishop(1995)}]{Bishop1995}
---. 1995, Neural Computation, 7, 108, \dodoi{10.1162/neco.1995.7.1.108}

\bibitem[{Blake {et~al.}(2012)Blake, Brough, Colless, Contreras, Couch, Croom,
  Croton, Davis, Drinkwater, Forster, Gilbank, Gladders, Glazebrook, Jelliffe,
  Jurek, Li, Madore, Martin, Pimbblet, Poole, Pracy, Sharp, Wisnioski, Woods,
  Wyder, \& Yee}]{Blake:2012}
Blake, C., Brough, S., Colless, M., {et~al.} 2012, Monthly Notices of the Royal
  Astronomical Society, 425, 405, \dodoi{10.1111/j.1365-2966.2012.21473.x}

\bibitem[{Bonassi \& West(2015)}]{Bonassi2015}
Bonassi, F.~V., \& West, M. 2015, Bayesian Analysis, 10, 171 ,
  \dodoi{10.1214/14-BA891}

\bibitem[{{Cai} {et~al.}(2016){Cai}, {Guo}, \& {Yang}}]{Cai2016}
{Cai}, R.-G., {Guo}, Z.-K., \& {Yang}, T. 2016, \prd, 93, 043517,
  \dodoi{10.1103/PhysRevD.93.043517}

\bibitem[{{Charnock} {et~al.}(2018){Charnock}, {Lavaux}, \&
  {Wandelt}}]{Charnock2018}
{Charnock}, T., {Lavaux}, G., \& {Wandelt}, B.~D. 2018, \prd, 97, 083004,
  \dodoi{10.1103/PhysRevD.97.083004}

\bibitem[{{Chen} {et~al.}(2023){Chen}, {Wang}, {Zhang}, \& {Zhang}}]{Chen2022}
{Chen}, J.-F., {Wang}, Y.-C., {Zhang}, T., \& {Zhang}, T.-J. 2023, \prd, 107,
  063517, \dodoi{10.1103/PhysRevD.107.063517}

\bibitem[{{Chen} {et~al.}(2024){Chen}, {Zhang}, {He}, {Zhang}, \&
  {Zhang}}]{Chen2024}
{Chen}, J.-f., {Zhang}, T.-J., {He}, P., {Zhang}, T., \& {Zhang}, J. 2024,
  arXiv e-prints, arXiv:2410.08369, \dodoi{10.48550/arXiv.2410.08369}

\bibitem[{{Christensen} {et~al.}(2001){Christensen}, {Meyer}, {Knox}, \&
  {Luey}}]{Christensen2001}
{Christensen}, N., {Meyer}, R., {Knox}, L., \& {Luey}, B. 2001, Classical and
  Quantum Gravity, 18, 2677, \dodoi{10.1088/0264-9381/18/14/306}

\bibitem[{{Chuang} {et~al.}(2017){Chuang}, {Pellejero-Ibanez},
  {Rodr{\'\i}guez-Torres}, {Ross}, {Zhao}, {Wang}, {Cuesta},
  {Rubi{\~n}o-Mart{\'\i}n}, {Prada}, {Alam}, {Beutler}, {Eisenstein},
  {Gil-Mar{\'\i}n}, {Grieb}, {Ho}, {Kitaura}, {Percival}, {Rossi},
  {Salazar-Albornoz}, {Samushia}, {S{\'a}nchez}, {Satpathy}, {Slosar},
  {Thomas}, {Tinker}, {Tojeiro}, {Vargas-Maga{\~n}a}, {Vazquez}, {Brownstein},
  {Nichol}, \& {Olmstead}}]{Chuang2017}
{Chuang}, C.-H., {Pellejero-Ibanez}, M., {Rodr{\'\i}guez-Torres}, S., {et~al.}
  2017, \mnras, 471, 2370, \dodoi{10.1093/mnras/stx1641}

\bibitem[{Cybenko(1989)}]{Cybenko1989}
Cybenko, G. 1989, Mathematics of Control, Signals and Systems, 2, 303,
  \dodoi{10.1007/BF02551274}

\bibitem[{Fisher(1922)}]{Fisher1921}
Fisher, R.~A. 1922, Journal of the Royal Statistical Society, 85, 87

\bibitem[{{Foreman-Mackey} {et~al.}(2013){Foreman-Mackey}, {Hogg}, {Lang}, \&
  {Goodman}}]{ForemanMackey2013}
{Foreman-Mackey}, D., {Hogg}, D.~W., {Lang}, D., \& {Goodman}, J. 2013, \pasp,
  125, 306, \dodoi{10.1086/670067}

\bibitem[{{Gazta{\~n}aga} {et~al.}(2009){Gazta{\~n}aga}, {Cabr{\'e}}, \&
  {Hui}}]{Gaztanaga2009}
{Gazta{\~n}aga}, E., {Cabr{\'e}}, A., \& {Hui}, L. 2009, \mnras, 399, 1663,
  \dodoi{10.1111/j.1365-2966.2009.15405.x}

\bibitem[{Gers {et~al.}(2000)Gers, Schmidhuber, \& Cummins}]{Gers2000}
Gers, F.~A., Schmidhuber, J., \& Cummins, F. 2000, Neural Computation, 12,
  2451, \dodoi{10.1162/089976600300015015}

\bibitem[{{G{\'o}mez-Valent} \& {Amendola}(2018)}]{GomezValent2018}
{G{\'o}mez-Valent}, A., \& {Amendola}, L. 2018, \jcap, 2018, 051,
  \dodoi{10.1088/1475-7516/2018/04/051}

\bibitem[{Graves {et~al.}(2013)Graves, Mohamed, \& Hinton}]{Graves2013}
Graves, A., Mohamed, A.-r., \& Hinton, G. 2013, in 2013 IEEE International
  Conference on Acoustics, Speech and Signal Processing, 6645--6649

\bibitem[{Hochreiter \& Schmidhuber(1997)}]{Hochreiter1997}
Hochreiter, S., \& Schmidhuber, J. 1997, Neural Computation, 9, 1735,
  \dodoi{10.1162/neco.1997.9.8.1735}

\bibitem[{Hochreiter \& Schmidhuber(1998)}]{Hochreiter1998}
---. 1998, International Journal of Bifurcation and Chaos, 8, 1071,
  \dodoi{10.1142/S0218488598000943}

\bibitem[{Hornik(1991)}]{Hornik1991}
Hornik, K. 1991, Neural Networks, 4, 251,
  \dodoi{https://doi.org/10.1016/0893-6080(91)90009-T}

\bibitem[{{Jesus} {et~al.}(2018){Jesus}, {Greg{\'o}rio}, {Andrade-Oliveira},
  {Valentim}, \& {Matos}}]{Jesus:2018}
{Jesus}, J.~F., {Greg{\'o}rio}, T.~M., {Andrade-Oliveira}, F., {Valentim}, R.,
  \& {Matos}, C.~A.~O. 2018, \mnras, 477, 2867, \dodoi{10.1093/mnras/sty813}

\bibitem[{Jiao {et~al.}(2023)Jiao, Borghi, Moresco, \& Zhang}]{Jiao2023}
Jiao, K., Borghi, N., Moresco, M., \& Zhang, T.-J. 2023, The Astrophysical
  Journal Supplement Series, 265, 48, \dodoi{10.3847/1538-4365/acbc77}

\bibitem[{{Jimenez} \& {Loeb}(2002)}]{Jimenez:2002}
{Jimenez}, R., \& {Loeb}, A. 2002, \apj, 573, 37, \dodoi{10.1086/340549}

\bibitem[{{Jimenez} {et~al.}(2003){Jimenez}, {Verde}, {Treu}, \&
  {Stern}}]{Jimenez:2003}
{Jimenez}, R., {Verde}, L., {Treu}, T., \& {Stern}, D. 2003, \apj, 593, 622,
  \dodoi{10.1086/376595}

\bibitem[{Jozefowicz {et~al.}(2015)Jozefowicz, Zaremba, \&
  Sutskever}]{Jozefowicz2015}
Jozefowicz, R., Zaremba, W., \& Sutskever, I. 2015, in Proceedings of Machine
  Learning Research, Vol.~37, Proceedings of the 32nd International Conference
  on Machine Learning, ed. F.~Bach \& D.~Blei (Lille, France: PMLR),
  2342--2350.
\newblock \url{https://proceedings.mlr.press/v37/jozefowicz15.html}

\bibitem[{LaValle {et~al.}(2004)LaValle, Branicky, \& Lindemann}]{lavalle2004}
LaValle, S.~M., Branicky, M.~S., \& Lindemann, S.~R. 2004, The International
  Journal of Robotics Research, 23, 673, \dodoi{10.1177/0278364904045481}

\bibitem[{{Lewis} \& {Bridle}(2002)}]{Lewis2002}
{Lewis}, A., \& {Bridle}, S. 2002, \prd, 66, 103511,
  \dodoi{10.1103/PhysRevD.66.103511}

\bibitem[{{Liao} {et~al.}(2019){Liao}, {Shafieloo}, {Keeley}, \&
  {Linder}}]{2019ApJ...886L..23L}
{Liao}, K., {Shafieloo}, A., {Keeley}, R.~E., \& {Linder}, E.~V. 2019, \apjl,
  886, L23, \dodoi{10.3847/2041-8213/ab5308}

\bibitem[{{Liu} {et~al.}(2022){Liu}, {Cao}, {Biesiada}, \&
  {Geng}}]{2022ApJ...939...37L}
{Liu}, T., {Cao}, S., {Biesiada}, M., \& {Geng}, S. 2022, \apj, 939, 37,
  \dodoi{10.3847/1538-4357/ac93f3}

\bibitem[{{Liu} {et~al.}(2025){Liu}, {Wang}, {Wu}, {Cao}, \&
  {Wang}}]{2025ApJ...981L..24L}
{Liu}, T., {Wang}, S., {Wu}, H., {Cao}, S., \& {Wang}, J. 2025, \apjl, 981,
  L24, \dodoi{10.3847/2041-8213/adb7de}

\bibitem[{Liu {et~al.}(2025)Liu, Wang, Vaidya, Ruehle, Halverson, Soljačić,
  Hou, \& Tegmark}]{Liu2024}
Liu, Z., Wang, Y., Vaidya, S., {et~al.} 2025, KAN: Kolmogorov-Arnold Networks.
\newblock \doarXiv{2404.19756}.
\newblock \url{https://arxiv.org/abs/2404.19756}

\bibitem[{{Ma} \& {Zhang}(2011)}]{Ma2011}
{Ma}, C., \& {Zhang}, T.-J. 2011, \apj, 730, 74,
  \dodoi{10.1088/0004-637X/730/2/74}

\bibitem[{Marjoram {et~al.}(2003)Marjoram, Molitor, Plagnol, \&
  Tavaré}]{Marjoram2003}
Marjoram, P., Molitor, J., Plagnol, V., \& Tavaré, S. 2003, Proceedings of the
  National Academy of Sciences, 100, 15324, \dodoi{10.1073/pnas.0306899100}

\bibitem[{{Mertens} {et~al.}(2018){Mertens}, {Ghosh}, \&
  {Koopmans}}]{2018MNRAS.478.3640M}
{Mertens}, F.~G., {Ghosh}, A., \& {Koopmans}, L.~V.~E. 2018, \mnras, 478, 3640,
  \dodoi{10.1093/mnras/sty1207}

\bibitem[{{Montiel} {et~al.}(2014){Montiel}, {Lazkoz}, {Sendra},
  {Escamilla-Rivera}, \& {Salzano}}]{Montiel2014}
{Montiel}, A., {Lazkoz}, R., {Sendra}, I., {Escamilla-Rivera}, C., \&
  {Salzano}, V. 2014, \prd, 89, 043007, \dodoi{10.1103/PhysRevD.89.043007}

\bibitem[{Moresco(2015)}]{Moresco:2015}
Moresco, M. 2015, Monthly Notices of the Royal Astronomical Society: Letters,
  450, L16, \dodoi{10.1093/mnrasl/slv037}

\bibitem[{{Moresco} {et~al.}(2012){Moresco}, {Verde}, {Pozzetti}, {Jimenez}, \&
  {Cimatti}}]{Moresco:2012}
{Moresco}, M., {Verde}, L., {Pozzetti}, L., {Jimenez}, R., \& {Cimatti}, A.
  2012, \jcap, 2012, 053, \dodoi{10.1088/1475-7516/2012/07/053}

\bibitem[{{Moresco} {et~al.}(2016){Moresco}, {Pozzetti}, {Cimatti}, {Jimenez},
  {Maraston}, {Verde}, {Thomas}, {Citro}, {Tojeiro}, \&
  {Wilkinson}}]{Moresco:2016}
{Moresco}, M., {Pozzetti}, L., {Cimatti}, A., {et~al.} 2016, \jcap, 2016, 014,
  \dodoi{10.1088/1475-7516/2016/05/014}

\bibitem[{{Pan} {et~al.}(2020){Pan}, {Liu}, {Forero-Romero}, {Sabiu}, {Li},
  {Miao}, \& {Li}}]{Pan2019}
{Pan}, S., {Liu}, M., {Forero-Romero}, J., {et~al.} 2020, Science China
  Physics, Mechanics, and Astronomy, 63, 110412,
  \dodoi{10.1007/s11433-020-1586-3}

\bibitem[{{Papamakarios} \& {Murray}(2016)}]{Papamakarios2016}
{Papamakarios}, G., \& {Murray}, I. 2016, arXiv e-prints, arXiv:1605.06376,
  \dodoi{10.48550/arXiv.1605.06376}

\bibitem[{{Papamakarios} {et~al.}(2017){Papamakarios}, {Pavlakou}, \&
  {Murray}}]{Papamakarios2017}
{Papamakarios}, G., {Pavlakou}, T., \& {Murray}, I. 2017, arXiv e-prints,
  arXiv:1705.07057, \dodoi{10.48550/arXiv.1705.07057}

\bibitem[{{Planck Collaboration} {et~al.}(2014){Planck Collaboration}, {Ade},
  {Aghanim}, {Armitage-Caplan}, {Arnaud}, {Ashdown}, {Atrio-Barandela},
  {Aumont}, {Baccigalupi}, {Banday}, {Barreiro}, {Bartlett}, {Bartolo},
  {Battaner}, {Battye}, {Benabed}, {Beno{\^\i}t}, {Benoit-L{\'e}vy}, {Bernard},
  {Bersanelli}, {Bielewicz}, {Bobin}, {Bock}, {Bonaldi}, {Bonavera}, {Bond},
  {Borrill}, {Bouchet}, {Bridges}, {Bucher}, {Burigana}, {Butler}, {Cardoso},
  {Catalano}, {Challinor}, {Chamballu}, {Chary}, {Chiang}, {Chiang},
  {Christensen}, {Church}, {Clements}, {Colombi}, {Colombo}, {Couchot},
  {Coulais}, {Crill}, {Cruz}, {Curto}, {Cuttaia}, {Danese}, {Davies}, {Davis},
  {de Bernardis}, {de Rosa}, {de Zotti}, {Delabrouille}, {Delouis},
  {D{\'e}sert}, {Diego}, {Dole}, {Donzelli}, {Dor{\'e}}, {Douspis}, {Ducout},
  {Dupac}, {Efstathiou}, {Elsner}, {En{\ss}lin}, {Eriksen}, {Fantaye},
  {Fergusson}, {Finelli}, {Forni}, {Frailis}, {Franceschi}, {Frommert},
  {Galeotta}, {Ganga}, {Giard}, {Giardino}, {Giraud-H{\'e}raud},
  {Gonz{\'a}lez-Nuevo}, {G{\'o}rski}, {Gratton}, {Gregorio}, {Gruppuso},
  {Hansen}, {Hansen}, {Hanson}, {Harrison}, {Helou}, {Henrot-Versill{\'e}},
  {Hern{\'a}ndez-Monteagudo}, {Herranz}, {Hildebrandt}, {Hivon}, {Hobson},
  {Holmes}, {Hornstrup}, {Hovest}, {Huffenberger}, {Jaffe}, {Jaffe}, {Jones},
  {Juvela}, {Keih{\"a}nen}, {Keskitalo}, {Kim}, {Kisner}, {Knoche}, {Knox},
  {Kunz}, {Kurki-Suonio}, {Lagache}, {L{\"a}hteenm{\"a}ki}, {Lamarre},
  {Lasenby}, {Laureijs}, {Lawrence}, {Leahy}, {Leonardi}, {Leroy},
  {Lesgourgues}, {Liguori}, {Lilje}, {Linden-V{\o}rnle}, {L{\'o}pez-Caniego},
  {Lubin}, {Mac{\'\i}as-P{\'e}rez}, {Maffei}, {Maino}, {Mandolesi}, {Mangilli},
  {Marinucci}, {Maris}, {Marshall}, {Martin}, {Mart{\'\i}nez-Gonz{\'a}lez},
  {Masi}, {Massardi}, {Matarrese}, {Matthai}, {Mazzotta}, {McEwen}, {Meinhold},
  {Melchiorri}, {Mendes}, {Mennella}, {Migliaccio}, {Mikkelsen}, {Mitra},
  {Miville-Desch{\^e}nes}, {Molinari}, {Moneti}, {Montier}, {Morgante},
  {Mortlock}, {Moss}, {Munshi}, {Murphy}, {Naselsky}, {Nati}, {Natoli},
  {Netterfield}, {N{\o}rgaard-Nielsen}, {Noviello}, {Novikov}, {Novikov},
  {Osborne}, {Oxborrow}, {Paci}, {Pagano}, {Pajot}, {Paoletti}, {Pasian},
  {Patanchon}, {Peiris}, {Perdereau}, {Perotto}, {Perrotta}, {Piacentini},
  {Piat}, {Pierpaoli}, {Pietrobon}, {Plaszczynski}, {Pogosyan},
  {Pointecouteau}, {Polenta}, {Ponthieu}, {Popa}, {Poutanen}, {Pratt},
  {Pr{\'e}zeau}, {Prunet}, {Puget}, {Rachen}, {Racine}, {R{\"a}th}, \&
  {Rebolo}}]{Aghanim2014}
{Planck Collaboration}, {Ade}, P.~A.~R., {Aghanim}, N., {et~al.} 2014, \aap,
  571, A23, \dodoi{10.1051/0004-6361/201321534}

\bibitem[{{Planck Collaboration} {et~al.}(2020){Planck Collaboration},
  {Aghanim}, {Akrami}, {Ashdown}, {Aumont}, {Baccigalupi}, {Ballardini},
  {Banday}, {Barreiro}, {Bartolo}, {Basak}, {Battye}, {Benabed}, {Bernard},
  {Bersanelli}, {Bielewicz}, {Bock}, {Bond}, {Borrill}, {Bouchet}, {Boulanger},
  {Bucher}, {Burigana}, {Butler}, {Calabrese}, {Cardoso}, {Carron},
  {Challinor}, {Chiang}, {Chluba}, {Colombo}, {Combet}, {Contreras}, {Crill},
  {Cuttaia}, {de Bernardis}, {de Zotti}, {Delabrouille}, {Delouis}, {Di
  Valentino}, {Diego}, {Dor{\'e}}, {Douspis}, {Ducout}, {Dupac}, {Dusini},
  {Efstathiou}, {Elsner}, {En{\ss}lin}, {Eriksen}, {Fantaye}, {Farhang},
  {Fergusson}, {Fernandez-Cobos}, {Finelli}, {Forastieri}, {Frailis},
  {Fraisse}, {Franceschi}, {Frolov}, {Galeotta}, {Galli}, {Ganga},
  {G{\'e}nova-Santos}, {Gerbino}, {Ghosh}, {Gonz{\'a}lez-Nuevo}, {G{\'o}rski},
  {Gratton}, {Gruppuso}, {Gudmundsson}, {Hamann}, {Handley}, {Hansen},
  {Herranz}, {Hildebrandt}, {Hivon}, {Huang}, {Jaffe}, {Jones}, {Karakci},
  {Keih{\"a}nen}, {Keskitalo}, {Kiiveri}, {Kim}, {Kisner}, {Knox},
  {Krachmalnicoff}, {Kunz}, {Kurki-Suonio}, {Lagache}, {Lamarre}, {Lasenby},
  {Lattanzi}, {Lawrence}, {Le Jeune}, {Lemos}, {Lesgourgues}, {Levrier},
  {Lewis}, {Liguori}, {Lilje}, {Lilley}, {Lindholm}, {L{\'o}pez-Caniego},
  {Lubin}, {Ma}, {Mac{\'\i}as-P{\'e}rez}, {Maggio}, {Maino}, {Mandolesi},
  {Mangilli}, {Marcos-Caballero}, {Maris}, {Martin}, {Martinelli},
  {Mart{\'\i}nez-Gonz{\'a}lez}, {Matarrese}, {Mauri}, {McEwen}, {Meinhold},
  {Melchiorri}, {Mennella}, {Migliaccio}, {Millea}, {Mitra},
  {Miville-Desch{\^e}nes}, {Molinari}, {Montier}, {Morgante}, {Moss}, {Natoli},
  {N{\o}rgaard-Nielsen}, {Pagano}, {Paoletti}, {Partridge}, {Patanchon},
  {Peiris}, {Perrotta}, {Pettorino}, {Piacentini}, {Polastri}, {Polenta},
  {Puget}, {Rachen}, {Reinecke}, {Remazeilles}, {Renzi}, {Rocha}, {Rosset},
  {Roudier}, {Rubi{\~n}o-Mart{\'\i}n}, {Ruiz-Granados}, {Salvati}, {Sandri},
  {Savelainen}, {Scott}, {Shellard}, {Sirignano}, {Sirri}, {Spencer},
  {Sunyaev}, {Suur-Uski}, {Tauber}, {Tavagnacco}, {Tenti}, {Toffolatti},
  {Tomasi}, {Trombetti}, {Valenziano}, {Valiviita}, {Van Tent}, {Vibert},
  {Vielva}, {Villa}, {Vittorio}, {Wandelt}, {Wehus}, {White}, {White},
  {Zacchei}, \& {Zonca}}]{Aghanim2018}
{Planck Collaboration}, {Aghanim}, N., {Akrami}, Y., {et~al.} 2020, \aap, 641,
  A6, \dodoi{10.1051/0004-6361/201833910}

\bibitem[{Ratsimbazafy {et~al.}(2017)Ratsimbazafy, Loubser, Crawford, Cress,
  Bassett, Nichol, \& Väisänen}]{Ratsimbazafy:2017}
Ratsimbazafy, A.~L., Loubser, S.~I., Crawford, S.~M., {et~al.} 2017, Monthly
  Notices of the Royal Astronomical Society, 467, 3239,
  \dodoi{10.1093/mnras/stx301}

\bibitem[{{Riess} {et~al.}(2007){Riess}, {Strolger}, {Casertano}, {Ferguson},
  {Mobasher}, {Gold}, {Challis}, {Filippenko}, {Jha}, {Li}, {Tonry}, {Foley},
  {Kirshner}, {Dickinson}, {MacDonald}, {Eisenstein}, {Livio}, {Younger}, {Xu},
  {Dahl{\'e}n}, \& {Stern}}]{2007ApJ...659...98R}
{Riess}, A.~G., {Strolger}, L.-G., {Casertano}, S., {et~al.} 2007, \apj, 659,
  98, \dodoi{10.1086/510378}

\bibitem[{{Samushia} {et~al.}(2013){Samushia}, {Reid}, {White}, {Percival},
  {Cuesta}, {Lombriser}, {Manera}, {Nichol}, {Schneider}, {Bizyaev},
  {Brewington}, {Malanushenko}, {Malanushenko}, {Oravetz}, {Pan}, {Simmons},
  {Shelden}, {Snedden}, {Tinker}, {Weaver}, {York}, \& {Zhao}}]{Samushia2013}
{Samushia}, L., {Reid}, B.~A., {White}, M., {et~al.} 2013, \mnras, 429, 1514,
  \dodoi{10.1093/mnras/sts443}

\bibitem[{{Scolnic} {et~al.}(2018){Scolnic}, {Jones}, {Rest}, {Pan},
  {Chornock}, {Foley}, {Huber}, {Kessler}, {Narayan}, {Riess}, {Rodney},
  {Berger}, {Brout}, {Challis}, {Drout}, {Finkbeiner}, {Lunnan}, {Kirshner},
  {Sanders}, {Schlafly}, {Smartt}, {Stubbs}, {Tonry}, {Wood-Vasey}, {Foley},
  {Hand}, {Johnson}, {Burgett}, {Chambers}, {Draper}, {Hodapp}, {Kaiser},
  {Kudritzki}, {Magnier}, {Metcalfe}, {Bresolin}, {Gall}, {Kotak}, {McCrum}, \&
  {Smith}}]{Scolnic2018}
{Scolnic}, D.~M., {Jones}, D.~O., {Rest}, A., {et~al.} 2018, \apj, 859, 101,
  \dodoi{10.3847/1538-4357/aab9bb}

\bibitem[{{Seikel} {et~al.}(2012){Seikel}, {Clarkson}, \& {Smith}}]{Seikel2012}
{Seikel}, M., {Clarkson}, C., \& {Smith}, M. 2012, \jcap, 2012, 036,
  \dodoi{10.1088/1475-7516/2012/06/036}

\bibitem[{{Shafieloo} {et~al.}(2012){Shafieloo}, {Kim}, \&
  {Linder}}]{2012PhRvD..85l3530S}
{Shafieloo}, A., {Kim}, A.~G., \& {Linder}, E.~V. 2012, \prd, 85, 123530,
  \dodoi{10.1103/PhysRevD.85.123530}

\bibitem[{{Simon} {et~al.}(2005){Simon}, {Verde}, \& {Jimenez}}]{Simon:2005}
{Simon}, J., {Verde}, L., \& {Jimenez}, R. 2005, \prd, 71, 123001,
  \dodoi{10.1103/PhysRevD.71.123001}

\bibitem[{{Stern} {et~al.}(2010){Stern}, {Jimenez}, {Verde}, {Kamionkowski}, \&
  {Stanford}}]{Stern:2010}
{Stern}, D., {Jimenez}, R., {Verde}, L., {Kamionkowski}, M., \& {Stanford},
  S.~A. 2010, \jcap, 2010, 008, \dodoi{10.1088/1475-7516/2010/02/008}

\bibitem[{{Sun} {et~al.}(2021){Sun}, {Jiao}, \& {Zhang}}]{Sun2021}
{Sun}, W., {Jiao}, K., \& {Zhang}, T.-J. 2021, \apj, 915, 123,
  \dodoi{10.3847/1538-4357/ac05b8}

\bibitem[{{Vaswani} {et~al.}(2017){Vaswani}, {Shazeer}, {Parmar}, {Uszkoreit},
  {Jones}, {Gomez}, {Kaiser}, \& {Polosukhin}}]{Vaswani2017}
{Vaswani}, A., {Shazeer}, N., {Parmar}, N., {et~al.} 2017, arXiv e-prints,
  arXiv:1706.03762, \dodoi{10.48550/arXiv.1706.03762}

\bibitem[{{Wang} {et~al.}(2023){Wang}, {Cheng}, {Ma}, {Xia}, {Abebe}, \&
  {Beesham}}]{Wang2023}
{Wang}, G.-J., {Cheng}, C., {Ma}, Y.-Z., {et~al.} 2023, \apjs, 268, 7,
  \dodoi{10.3847/1538-4365/ace113}

\bibitem[{{Wang} {et~al.}(2020{\natexlab{a}}){Wang}, {Li}, \&
  {Xia}}]{Wang2020ecopann}
{Wang}, G.-J., {Li}, S.-Y., \& {Xia}, J.-Q. 2020{\natexlab{a}}, \apjs, 249, 25,
  \dodoi{10.3847/1538-4365/aba190}

\bibitem[{{Wang} {et~al.}(2020{\natexlab{b}}){Wang}, {Ma}, {Li}, \&
  {Xia}}]{Wang2020reann}
{Wang}, G.-J., {Ma}, X.-J., {Li}, S.-Y., \& {Xia}, J.-Q. 2020{\natexlab{b}},
  \apjs, 246, 13, \dodoi{10.3847/1538-4365/ab620b}

\bibitem[{{Wang} {et~al.}(2021){Wang}, {Xie}, {Zhang}, {Huang}, {Zhang}, \&
  {Liu}}]{WangY2020}
{Wang}, Y.-C., {Xie}, Y.-B., {Zhang}, T.-J., {et~al.} 2021, \apjs, 254, 43,
  \dodoi{10.3847/1538-4365/abf8aa}

\bibitem[{{Weyant} {et~al.}(2013){Weyant}, {Schafer}, \&
  {Wood-Vasey}}]{Weyant2013}
{Weyant}, A., {Schafer}, C., \& {Wood-Vasey}, W.~M. 2013, \apj, 764, 116,
  \dodoi{10.1088/0004-637X/764/2/116}

\bibitem[{{Zhang} {et~al.}(2014){Zhang}, {Zhang}, {Yuan}, {Liu}, {Zhang}, \&
  {Sun}}]{Zhang:2014}
{Zhang}, C., {Zhang}, H., {Yuan}, S., {et~al.} 2014, Research in Astronomy and
  Astrophysics, 14, 1221, \dodoi{10.1088/1674-4527/14/10/002}

\bibitem[{{Zhang} {et~al.}(2024){Zhang}, {Hu}, {Jiao}, {Wang}, {Xie}, {Yu},
  {Zhao}, \& {Zhang}}]{Zhang2024}
{Zhang}, J.-C., {Hu}, Y., {Jiao}, K., {et~al.} 2024, \apjs, 270, 23,
  \dodoi{10.3847/1538-4365/ad0f1e}

\bibitem[{{Zhou} \& {Li}(2019)}]{Zhou:2019}
{Zhou}, H., \& {Li}, Z. 2019, Chinese Physics C, 43, 035103,
  \dodoi{10.1088/1674-1137/43/3/035103}

\end{thebibliography}
\bibliographystyle{aasjournal}

\end{document}